\documentclass[conference]{IEEEtran}

\usepackage{caption}
\usepackage{color}
\usepackage{graphicx}
\usepackage{amsmath, amssymb, amsfonts, amsthm}
\usepackage{fdsymbol}
\usepackage{nicefrac}
\usepackage{prooftree}
\usepackage{cite}
\usepackage{spi}
\usepackage[draft]{fixme}
\usepackage{cancel}
\usepackage{url}
\usepackage{mathtools}
\usepackage{eucal}[mathcal]
\usepackage{yfonts}
\usepackage{listings}
\usepackage[shortlabels]{enumitem}
\usepackage[normalem]{ulem}
\usepackage{xcolor}
\def\BibTeX{{\rm B\kern-.05em{\sc i\kern-.025em b}\kern-.08em
		T\kern-.1667em\lower.7ex\hbox{E}\kern-.125emX}}

\lstdefinestyle{stl}{
	basicstyle=\ttfamily
}
\newtheorem{definition}{Definition}
\newtheorem{theorem}{Theorem}

\newtheorem{lemma}[theorem]{Lemma}

\newcommand{\cch}{\textit{card}}
\newcommand{\tch}{\textit{term}}
\newcommand{\System}{\textit{SYS}}
\newcommand{\nf}[1]{\mathopen{#1}\downharpoonright}
\newcommand{\pspec}{\nw c. \nw ch. \cout{card}{ch}.\fix{\card}}
\newcommand{\pimpl}{\nw c. \bang \nw ch. \cout{card}{ch}.\fix{\card}}
\newcommand\certificate{\mathopen\textit{cert}}
\newcommand\fresh{\#}
\renewcommand{\lts}[1]{\xrightarrow{#1}}
\newcommand{\diam}[1]{\mathopen{\big\langle #1 \big\rangle}}
\newcommand{\bdh}[1]{{#1}_{\text{rfc}}}
\newcommand{\spec}[1]{\textit{#1}_{\text{spec}}}
\newcommand{\impl}[1]{\textit{#1}_{\text{impl}}}
\newcommand{\fix}[1]{{#1}_{\text{upd}}}
\newcommand{\smult}[2]{\mathopen\phi\left(#1, #2\right)}
\renewcommand{\pk}[1]{\smult{#1}{\gen}}
\newcommand{\gen}{\mathtt{g}}

\newcommand{\ack}{\mathopen\texttt{auth}}

\newcommand{\m}[1]{\mathcal{#1}}

\newcommand{\checksig}[2]{\mathopen\texttt{check}\left(#2, #1\right)}
\newcommand{\pks}[1]{\mathopen\texttt{pk}\left(#1\right)}
\newcommand{\key}[1]{\mathopen\textit{k}_{#1}}
\newcommand{\lett}{\texttt{let}}
\newcommand{\nw}{\mathopen\nu}
\newcommand{\card}{\mathopen C}
\newcommand{\terminal}{\mathopen T}
\newcommand{\quotes}[1]{``#1''}
\newcommand{\sig}[2]{\mathopen\texttt{sig}\left(#2, #1\right)}

\newcommand{\proj}[1]{\mathopen\texttt{fst}\left(#1\right)}
\newcommand{\projj}[1]{\mathopen\texttt{snd}\left(#1\right)}
\newcommand{\mult}[2]{\ensuremath{#1 \cdot #2}}
\newcommand{\match}[1]{\mathopen{\texttt{if} \, #1 \, \texttt{then} \,}}
\newcommand{\angles}[1]{\left<{#1}\right>}
\newcommand{\rounds}[1]{\left({#1}\right)}
\renewcommand{\dec}[2]{\mathopen\texttt{dec}\left(#2, #1\right)}
\newcommand*\squeezespaces[1]{
	\thickmuskip=\scalemuskip{\thickmuskip}{#1}
	\medmuskip=\scalemuskip{\medmuskip}{#1}
	\thinmuskip=\scalemuskip{\thinmuskip}{#1}
	\nulldelimiterspace=#1\nulldelimiterspace
	\scriptspace=#1\scriptspace
}
\newcommand*\scalemuskip[2]{%
	\muexpr #1*\numexpr\dimexpr#2pt\relax\relax/65536\relax
}
\DeclareTextFontCommand{\code}{\ttfamily}

\IEEEoverridecommandlockouts
\begin{document}
	
	\title{
		Unlinkability of an Improved Key Agreement Protocol for EMV 2nd Gen Payments
	}
	\author{\IEEEauthorblockN{Ross Horne}
		\IEEEauthorblockA{\textit{Department of Computer Science} \\
			\textit{University of Luxembourg}\\
			Esch-sur-Alzette, Luxembourg \\
			ross.horne@uni.lu}
		\and
		\IEEEauthorblockN{Sjouke Mauw}
		\IEEEauthorblockA{\textit{Department of Computer Science} \\
			\textit{University of Luxembourg}\\
			Esch-sur-Alzette, Luxembourg  \\
			sjouke.mauw@uni.lu}
		\and
		\IEEEauthorblockN{Semen Yurkov$^\dagger$}
		\IEEEauthorblockA{\textit{Department of Computer Science} \\
			\textit{University of Luxembourg}\\
			Esch-sur-Alzette, Luxembourg  \\
			semen.yurkov@uni.lu}
		\thanks{$^{\dagger}$Semen Yurkov is supported by the Luxembourg National Research Fund through grant PRIDE15/10621687/SPsquared.}
	}
	
	\maketitle
	
	\begin{abstract}
		To address known privacy problems with the EMV standard, EMVCo have proposed a \emph{Blinded Diffie-Hellman} key establishment protocol, which is intended to be part of a future 2nd Gen EMV protocol.
		We point out that active attackers were not previously accounted for in the privacy requirements of this proposal protocol, and demonstrate that an active attacker can compromise unlinkability within a distance of 100cm.
		Here, we adopt a strong definition of unlinkability that does account for
		active attackers and propose an enhancement of the protocol proposed by EMVCo. 
		We prove that our protocol does satisfy strong unlinkability, while preserving authentication.
	\end{abstract}
	
	
	\section{Introduction}
	The majority of payment cards and terminals use the EMV standard~\cite{emv}, the set of protocols developed by the union of
	payment processing companies Europay, Mastercard and Visa to execute financial operations. The initial purpose of EMV, introduced in 1996, was to support the replacement of mag-stripe cards with integrated circuit cards that are harder to copy. The EMV standard now supports contactless cards, that require no cardholder action to be involved in the EMV session with any capable device. The nature of contactless cards allows an active attacker to easily interact with the card without the cardholder realising, making privacy properties harder to enforce.
	
	In this paper, we address privacy vulnerabilities in payment cards with a particular focus on the \emph{unlinkability} of payments. 
	The EMV standard trivially does not satisfy privacy properties such as anonymity and unlinkability. This is due to the current EMV standard transferring the card number in cleartext during a transaction. Hence transaction data allows us to link transactions
	made with the same card and effortlessly track cardholders. The fact that no actual payments need to be made eases the task of the adversary when tracking a contactless card as it is ready to present its identity to any device. 
	
	In 2011 EMVCo launched the development of the new version of the standard, the EMV 2nd Gen, where the card should be protected against eavesdropping. To facilitate this, EMVCo proposed the use of \emph{secret channels}. A secret channel is a symmetric key that the card and the terminal establish at the start of each session and use to encrypt further communications. A channel establishment procedure is based on Diffie-Hellman key agreement with a twist: the card uses a freshly blinded static certified public key instead of an ephemeral public key. Hence, the name of the proposed protocol, \emph{Blinded Diffie-Hellman} (BDH)~\cite{rfc}.
	
	The BDH protocol is meant to satisfy the official requirements for channel establishment from the architecture overview of the EMV 2nd Gen~\cite{overview}:
	
	\begin{itemize}
		\item Use elliptic-curve based cryptography (ECC). 
		\item Computational resources of the card are respected.
		\item An attacker who passively eavesdrops on communications cannot identify a particular card.
	\end{itemize}

	Several authors published a security proof for the Blinded Diffie-Hellman protocol~\cite{brzuska2013analysis, guo2014security} and established that a passive eavesdropper, that only listens to transmitted messages, cannot reidentify a card, therefore BDH satisfies the above requirements. Brzuska, Smart, Warinschi, and Watson~\cite{brzuska2013analysis} named this property of BDH \quotes{external unlinkability}.

	We discuss a potential strengthening of the requirements for BDH listed above, i.e. we lift the limitation of
	attackers being passive. In the context of contactless payments such requirement is a realistic one, since it is easy for an attacker to initiate sessions with contactless cards using devices, such as smartphones, that need not be official terminals. In fact, different capabilities must be considered in a wireless environment depending on their distance from the card. It is difficult to perform an eavesdropping attack outside of the approximately 20m radius~\cite{pfeiffer2012finkenzeller}. However, successful attacks executed by \emph{a passive attacker} are reported within the range between 20m and 100cm~\cite{novotny2008guerrieri,
		engelhardt2013pfeiffer} and, with the right equipment, within 100cm
	\emph{an active attacker} can power up the card and start
	communication~\cite{habraken2015dolron}. A close
	active attacker is a real threat to the privacy of anyone having a card in their pocket, since the distance of 100cm is easily achievable, e.g. at doorways or checkouts. 
	
	Unsurprisingly, the proposed BDH protocol is no longer secure in such strictly stronger threat model. To accommodate our strengthened threat model we consider an enhancement of the proposed protocol that is unlinkable and untraceable. The definition of unlinkability we propose is formalised as a process equivalence problem in the applied $\pi$-calculus and accommodates active attackers. Our enhancement of BDH uses a generic \emph{anonymous credentials} scheme to hide the card's identity. At least one anonymous credential scheme, Verheul certificates, is known to respect the limited computing power of smart cards making our upgrade minor.

	To define unlinkability in the context of EMV payments, we are building on the state-of-the-art bisimilarity-based approach developed by Horne and Mauw~\cite{horne2021mauw}. The use of quasi-open bisimilarity~\cite{horne2021quasi} in our new definition has the following two reasons behind it. 
	\begin{itemize}
		\item It helps to reduce drastically the amount of work needed for
		verification: being a congruence relation, quasi-open bisimilarity
		allows us to take into account cards only, since not having a common secret between cards and terminals is a part of the philosophy of EMV.
		\item It ensures that our results hold also in weaker models like trace equivalence: being a bisimilarity, quasi-open bisimilarity is a strictly finer equivalence. 
	\end{itemize}
	
	\noindent
	The contributions of the work are as follows.
	
	\begin{itemize}
		\item A new definition for unlinkability suitable for EMV payments that accounts for active attackers.
		\item An attack invalidating our strong
		unlinkability goal on Blinded Diffie-Hellman, in the form initially
		proposed by EMVCo described by a modal logic formula.
		\item An improved proposal for the Blinded Diffie-Hellman protocol by integrating blind certificates. 
		\item The proof of the unlinkability of our improved BDH.
		\item A discussion on the unlinkable, in our stronger sense, EMV transactions.
	\end{itemize}
	
	The paper is organised as follows. 
	Section~\ref{sec_rel} is devoted to previous work on EMV security and privacy issues. In Section~\ref{sec_bdh}, we present the original Blinded Diffie-Hellman and illustrate why there are attacks on unlinkability in the presence of an active attacker. 
	In Section~\ref{sec_pi}, we provide background on the applied $\pi$-calculus.
	Sections~\ref{sec_unlink} and~\ref{sec_making} contain the main contributions of the paper: a new definition of unlinkability for EMV payments, an enhanced version of BDH that includes blinded certificates, and a proof of the unlinkability of this enhanced version.
	Section~\ref{sec_auth} confirms that authentication properties are preserved by our enhanced protocol. 
	In Section~\ref{sec_discussion} we discuss unlinkable transactions with respect to our strong threat model and explain that it would be impossible without touching the fundamentals of the current EMV infrastructure, since the account number would eventually be received by the dishonest terminal.
	Section~\ref{sec_concl} concludes the paper and presents directions for future research.

	\section{Related work}\label{sec_rel}
	
	Much related work on EMV is concerned with authentication and secrecy problems essential for avoiding fraudulent payments.
	The recent works of Basin, Sasse, and Toro-Pozo~\cite{jorge2020emv, jorge2021emv}
	contain an overview of attacks on EMV that can lead to fraudulent transactions,
	e.g. criminals can make high-value purchases 
	using a contactless Visa 
	card without knowing the PIN.
	Contactless specific \emph{relay} attacks may be mitigated by using 
	distance-bounding techniques~\cite{mauw2018db}, 
	for instance Chothia, de Ruiter and Smyth verified Mastercard's RRP protocol ~\cite{chothia2018ruiter}, Boureanu et al.\ analyse relay-resistance EMV-based protocols in the presence of rogue readers ~\cite{boureanu2020contactless}, and Radu et al. combine man-in-the-middle replay and relay attacks to bypass the Apple Pay (working with Visa card) lock screen~\cite{radu2022practical}. 
	In a \emph{skimming attack}, an attacker secretly activates a contactless card and communicates with it. A skimming attack may be a part of a relay attack and serves as the basis of the attack on the unlinkability of BDH we present in this paper. Habraken et al. constructed an antenna in the form of a gate of up to 100cm width that can power the card and communicate with it~\cite{habraken2015dolron}. For a passive counterpart of skimming, an \emph{eavesdropping attack}, Engelhardt et al. achieved the distance of 18m~\cite{engelhardt2013pfeiffer}.
	
	To enhance privacy in EMV it is natural to consider anonymous credentials systems, since they allow credentials to be verified without disclosing the identity of a cardholder; although such mechanisms have not been explored in the context of EMV payments. Idemix~\cite{camenisch2001efficient} and
	U-Prove~\cite{brands2000rethinking} are general-purpose examples of
	such systems. However, they barely fit the context of this paper since
	Idemix requires a large key size, therefore implementation on
	smart-cards is rather slow~\cite{tews2009performance} and it is
	straightforward to link transactions in U-Prove when the same credentials
	(in our case, the card's identity) are used twice. A more suitable
	anonymous credential system 
	is the self-blindable attribute
	certificates due to Verheul~\cite{verheul2001self}. Verheul certificates use
	elliptic curve cryptography, aligning with stated requirements of EMV 2nd
	Gen, and have been demonstrated to be efficiently implementable on smart cards~\cite{batina2010developing}.
	
	Arapinis et al.~\cite{arapinis2010analysing} proposed to express unlinkability as an equivalence problem; specifically, they defined \emph{strong unlinkability} using bisimulation. Horne and Mauw~\cite{horne2021mauw} propose a new scheme by adding session channels to the model and thoroughly study the advantages of the bisimilarity approach. We partially employ the formulation of the $\pi$-calculus presented in their work. Hirschi, Baelde, and Delaune~\cite{baelde19amethod} weakened the definition from~\cite{arapinis2010analysing} by redefining unlinkability as a trace equivalence problem for which they develop tool support for obtaining proofs of unlinkability. Using trace equivalence, however, may lead to missing attacks as pointed out in~\cite{filimonov2019breaking} where Filimonov et al. study ePassport protocols and revisit bisimilarity-based strong unlinkability definitions. There is an ongoing debate~\cite{baelde19amethod, filimonov2019breaking} on the benefits of each equivalence, but in this work either is appropriate since we prove properties in the strongest of these models and find attacks in the weakest.
	
	Finally, we mention works on symbolic methods for analysing
	Diffie-Hellman (DH) groups. The general case requires both exponentiation and the group 
	operation to be modelled and a straightforward
	approach may lead to the unification problem in a
	field~\cite{schmidt2012formal} which is undecidable. Tools like 
	Tamarin~\cite{meier2013tamarin} or ProVerif~\cite{kusters2009dhexp} use prime 
	order group abstractions to facilitate verification. Cremers and Jackson investigate in detail 
	the subtleties of modelling DH groups in automated tools and propose
	improved models in~\cite{cremers2019prime}.
	
	\section{Blinded Diffie-Hellman and external unlinkability}\label{sec_bdh}
	In this motivating section, we introduce the Blinded Diffie-Hellman protocol from the original EMVCo request for comments~\cite{rfc} and highlight its unlinkability issues.
	
	\subsection{The Blinded Diffie-Hellman protocol}\label{sec_bdh_asis}
	
	To present the BDH protocol we define the syntax of messages in Fig.~\ref{fig_syntax}.
	\begin{figure}[h]
		\[\arraycolsep=3.5pt
		\begin{gathered}
			\begin{array}{c}
				\begin{array}{rlr}
					M, N \Coloneqq
					& \gen & \mbox{DH group generator (constant)} \\
					\small|& x & \mbox{variable} \\
					\small|& \mult{M}{N} & \mbox{multiplication} \\
					\small|& \smult{M}{N} & \mbox{scalar multiplication} \\
					\small|& \pks{M} & \mbox{public key} \\
					\small|& \sig{N}{M} & \mbox{signature} \\
					\small|& \hash{M} & \mbox{hash (for key derivation)} \\
					\small|& \pair{M}{N} & \mbox{pair} \\
					\small|& \enc{M}{N} & \mbox{symmetric encryption} \\
					\small|& \checksig{N}{M} & \mbox{check signature} \\
					\small|& \proj{M} & \mbox{get first} \\
					\small|& \projj{M} & \mbox{get second} \\
					\small|& \dec{N}{M} & \mbox{symmetric decryption} \\
					\small|& \ack & \mbox{authenticate}\\
				\end{array} 
			\end{array}
		\end{gathered}
		\]
		\caption{Blinded Diffie-Hellman syntax.} 
		\label{fig_syntax}
	\end{figure}
	
	The syntax for messages includes abstractions for the arithmetic operations on elliptic curves that protocols in this paper employ, enabling us to represent protocols symbolically.
	We leave the cryptographic details for multiplication, scalar multiplication and public key operations together with ECC domain parameters as a footnote\footnote{\scriptsize
		The public parameters are as follows: a finite field $\mathbb{F}_p$; a Diffie-Hellman group $G$, defined over an elliptic curve $E(\mathbb{F}_p)$; the (prime) order $q$ of $G$; the generator $\gen \in G$; the key-derivation function $h$; the public key of the payment system $\pks{s}$ for the certificate verification. We employ (left) group action notation $\phi \colon \mathbb{F}_q^\times \times G \rightarrow G$ for group operation: we write $\smult{r}{Q}$ for the element $Q$ added with itself $r$ times and call $\phi$ \emph{scalar multiplication}. The symbol~$\cdot$ denotes multiplication between two scalars (field elements). All freshly generated values are picked uniformly at random from $\mathbb{F}_q$. The secret key $k$ is an element of $\mathbb{F}_q$ and the corresponding public key is of the form $\smult{k}{\gen}$. Blinding of the element $Q$ uses a fresh scalar $a$ and internally works as a scalar multiplication: $\smult{a}{Q}$. 
	}. The exact signing mechanism modelled by $\sig{N}{M}$ is not specified by EMVCo in the proposal~\cite{rfc}. Hash, pair and encryption are standard and $\ack$ is a message that upon being output indicates that the terminal believes it has authenticated the card. The authentication property is explained in Section~\ref{sec_auth}.
	
	The equational theory $E_0$ axiomatising the properties of the cryptographic functions is given in Fig.~\ref{fig_eqt} The first three equations capture the interaction between field arithmetic and scalar multiplication followed by standard destructors: projections, decryption, and signature check. Notice that we model signature verification in a manner that is standard when symbolically verifying protocols: the signature is verified iff the message is successfully extracted by applying $\texttt{check}$ from $\sig{M}{K}$ using the corresponding public key $\pks{K}$.
	\begin{figure}[h]
		\[
		\begin{gathered}
			\begin{array}{c}
				\begin{array}{ll}
					\mult{M}{N} =_{E_0} \mult{N}{M} \\[1pt]
					\mult{\rounds{\mult{M}{N}}}{K} =_{E_0} \mult{M}{\rounds{\mult{N}{K}}} \\[1pt]
					\smult{\mult{M}{N}}{K} =_{E_0} \smult{M}{\smult{N}{K}} \\[1pt]
					\proj{\pair{M}{N}} =_{E_0} M \\[1pt]
					\projj{\pair{M}{N}} =_{E_0} N \\[1pt]
					\dec{K}{\enc{M}{K}} =_{E_0} M \\[1pt]
					\checksig{\pks{K}}{\sig{K}{M}} =_{E_0} M \\[1pt]
				\end{array}
			\end{array}
		\end{gathered}
		\]
		\caption{Equational theory $E_0$ for the Blinded Diffie-Hellman protocol.} 
		\label{fig_eqt}
	\end{figure}
	
\begin{figure}[h]
		\includegraphics[width=\linewidth]{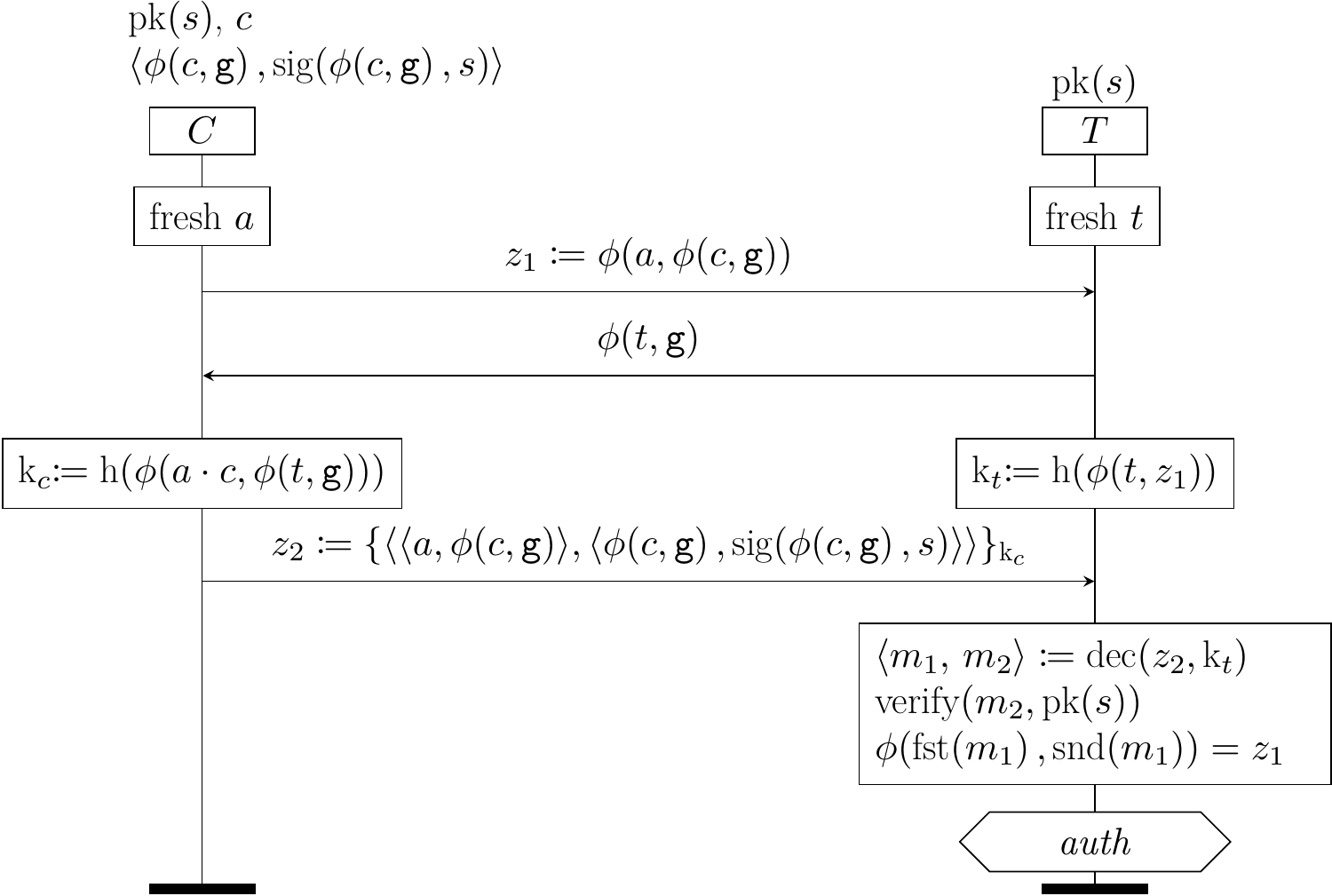}
		\centering
		\caption{EMV 2nd Gen key establishment.} 
		\label{fig_orig}
	\end{figure}

	The Blinded Diffie-Hellman protocol is presented in Fig.~\ref{fig_orig}.
	There are two honest agents in the system that 
	participate in the execution of the protocol: the card $C$ and the terminal $T$. The payment system holds a secret key $s$ and acts as a certification authority. The private key $c$, the public key $\pk{c}$ and the certificate $\pair{\pk{c}}{\sig{s}{\pk{c}}}$ are permanently embedded in the card when it is manufactured. The card can only be issued by the bank in cooperation with payment systems like Amex, Visa, etc. The terminal, in contrast to the card, can be manufactured by anyone. To verify the legitimacy of the card, the terminal uses a public key of the payment system $\pks{s}$ that is available on the system's website.
	
	The card starts the communication by sending its public key $\pk{c}$ blinded with a fresh scalar $a$ to the terminal. In response, the terminal sends ephemeral public key $\pk{t}$ to the card. This is enough to establish a common secret key $\key{c}=\key{t}$. The card uses this key to encrypt the authentication data: blinding scalar $a$, static public key $\pk{c}$, and the certificate $\langle \pk{c}, \sig{s}{\pk{c}}\rangle\rangle$. Finally, the terminal verifies the received certificate by checking the signature against the public key of the payment system $\pks{s}$, checks that $\pk{c}$ blinded with $a$ coincides with the first message $z_1$ received from the card. 
	Upon success, the terminal authenticates the card and is ready to continue with the transaction on the encrypted channel.

	\subsection{Blinded Diffie-Hellman and active attackers}\label{sec_notunlinkable}
	
	In order to verify that blinding the card's public key protects against eavesdroppers external to the execution, the property of \emph{external unlinkability} was introduced~\cite{brzuska2013analysis}.
	In an externally unlinkable payment system, an attacker observing
	a message exchange between a card and a terminal cannot link that card's current session with a previous session from the same card. 
	
	In the real world, anyone could build a device imitating the terminal, for instance, an app on a smartphone supporting NFC or a skimming gate~\cite{habraken2015dolron}.
	Such a device need not be certified or connected to any bank. Taking this into account, there is a straightforward attack on the BDH protocol (Fig.~\ref{fig_orig}) in the presence of malicious terminals:
	\begin{enumerate}
		\item 
		A malicious terminal establishes a key with an honest card, then successfully decrypts the message $z_2$ and obtains the card's public key $\phi(c,\gen)$.
		\item Another terminal operated by the attacker runs a new session with the same card to obtain again the card's public key $\phi(c,\gen)$; and hence recognises the card.
	\end{enumerate}
	
	This attack however would not be considered to be an attack on external unlinkability, due to the fact that, at the second step, 
	the attacker actively starts comunicating with the card. Since it is easy to activate a contactless card, e.g. while the card is in the wallet, external unlinkability is too weak. This compels us to adopt a stronger notion of unlinkability which can be used to discover the above attack formally.
	
	The above attack suggests that any network of malicious powerful terminal-like devices unrelated to any payment system may track selected contactless cards in real-time without the cardholder being aware simply by starting sessions with the card in the cardholder's pocket. 
	Thus we propose to view unlinkability as a \emph{property of the card}
	in a hostile environment that should hold with or without the presence
	of honest terminals. The attack also highlights why the BDH protocol is not unlinkable in the presence of active attackers -- the ability of the terminal to obtain the card's public key which serves as the card's identity.
	
	To address the unlinkability vulnerability highlighted above, we propose to modify the Blinded Diffie-Hellman protocol 
	in such a way that the signature may also be blinded and hence the public key need never be revealed to a terminal in order to check the signature.
	We will present and verify our enhanced version of BDH in Section~\ref{sec_making}. However, first, we dedicate the next two sections to the machinery required to formally specify and verify that our proposal is unlinkable.

	\section{Applied $\pi$-calculus and quasi-open bisimilarity}\label{sec_pi}
	This section contains background on a state-of-the-art formulation of the applied $\pi$-calculus~\cite{abadi2001mobile}, a language for modelling concurrent processes and their interactions. The calculus is presented in a reduced form that is just enough for the purpose of the paper. We start with the syntax and move towards the definition of an equivalence relation on processes that we use to express the unlinkability definition in Section~\ref{sec_unlink}.
	\subsection{Syntax, notation, conventions}\label{sec_syntax}
	The syntax of processes is presented in Fig.~\ref{fig_pisyntax}.
	\begin{figure}[h]
		\[
		\begin{gathered}
			\begin{array}{rlr}
				P, Q \Coloneqq& 0 & \mbox{deadlock} \\
				\small|& \cout{M}{N}.P & \mbox{send} \\ 
				\small|& \cin{M}{y}.P & \mbox{receive} \\
				\small|& \mathopen\nu x. P & \mbox{new} \\
				\small|& P \cpar Q & \mbox{parallel} \\
				\small|& \bang P & \mbox{replication} \\ 
				\small|& \match{M = N}P & \mbox{match} \\
			\end{array}
		\end{gathered}
		\]
		\caption{A syntax for processes in applied $\pi$-calculus processes.}
		\label{fig_pisyntax}
	\end{figure}
	
	Processes are used to capture the behaviour of a system, and, in
	particular, a behaviour of honest parties during the execution of a
	protocol. Processes can output and consume messages. To do that they
	use \emph{channels}, e.g. $\cout{M}{N}$ means that the message $N$ is
	sent out on the channel $M$. Messages can be defined with respect to
	any message language (e.g. in Fig.~\ref{fig_syntax}) subject to any equational theory (e.g. in Fig.~\ref{fig_eqt}). We write $M =_{E} N$ for equality modulo an equational theory $E$. 
	
	Variables in processes may be \emph{bound} by new name binders or inputs: specifically $\nu x.P$ and $M(x).P$ bind $x$ in the scope $P$. In other words, the variable $x$ becomes local to the process $P$. If a variable is not bound, it is a \emph{free} variable. We denote by $\fv{T}$ the set of free variables in a process or a message term $T$.
	
	The processes $P$ and $Q$ in $P \cpar Q$ run concurrently. The replication $\bang P$ is an infinite parallel composition of $P$ with itself. Finally, the process $\match{M = N}P$ can behave as $P$ whenever $M \mathrel{=_{E}} N$.
	
	A \emph{substitution} is a function from a finite set of variables to message terms. We use vector notation to indicate the list of variables $\vec{x}$ or messages $\vec{M}$. Whenever $\vec{x}$ is involved in set-theoretic operations we treat $\vec{x}$ as the set of variables in $\vec{x}$. We use $\sigma$, $\rho$ and $\theta$ to refer to substitutions and write $x\sigma$ for $\sigma$ applied to the variable $x$. The result of applying the substitution $\sigma$ to the process $P$ is the replacement of any free occurrence of $x$ in $P$ with $x\sigma$. We write $P\sigma$ for the resulting process. When $\sigma$ is given explicitly, we write $\sigma = \sub{\vec{x}}{\vec{M}}$. Substitutions must avoid capture of bound variables: if a bound variable $x$ in the process $P$ occurs in the range of $\sigma$, it must be renamed to avoid a name clash. The renaming of bound variables is a standard operation in the $\pi$-calculus known as $\alpha$-conversion~\cite{sangiorgi01walker} and we always consider processes up to $\alpha$-conversion. For instance, to compute $\cin{a}{x}.\cout{a}{\enc{\pair{x}{y}}{k}}\sub{y}{\hash{x}}$ we apply $\alpha$-conversion first and get $\cin{a}{z}.\cout{a}{\enc{\pair{z}{y}}{k}}\sub{y}{\hash{x}}$, where $z$ is chosen 
	fresh for $\hash{x}$ and $\cout{a}{\enc{\pair{x}{y}}{k}}$, i.e., $z \not\in \{a,x,y,k\}$, and then apply the substitution to obtain the result $\cin{a}{z}.\cout{a}{\enc{\pair{z}{\hash{x}}}{k}}$.
	
	We generalise the concept of a variable not belonging to some set of variables in the following definition.
	\begin{definition}\emph{(fresh, \fresh)}\label{def_fresh}
		The set of variables $\vec{x}$ is fresh for the set of variables $\vec{y}$ if $\vec{x} \cap \vec{y} = \emptyset$; $\vec{x}$ is fresh for a term $P$ if $\vec{x}$ is fresh for $\fv{P}$; $\vec{x}$ is fresh for a substitution $\sigma$ whenever $\vec{x}$ is fresh for $\dom{\sigma}$ and fresh for $\fv{y\sigma}$ for any $y$ fresh for $\vec{x}$. Notation: $\vec{x}$ \emph{\fresh{}} $\vec{y}$, $\vec{x}$ \emph{\fresh{}} $P$, $\vec{x}$ \emph{\fresh{}} $\sigma$.
	\end{definition}
	That is, fresh variables never appear in the set of free variables or
	the domain and the range of the substitution.
	
	Throughout the paper we use several conventions. We do not distinguish between $\nw x_1. \nw x_2.P$ and $\nw x_1. (\nw x_2.P)$ and typically write $\nw x_1, x_2.P$. The symbol $\triangleeq$ is used to define a process. For readability purposes we introduce the following abbreviations.
	\[
	\begin{array}{l}
		\lett\,x \coloneqq M \ \texttt{in} \ P \ \triangleq \ P\sub{x}{M}
		\\[5pt]
		\lett \pair{x_1}{x_2}=M \ \texttt{in} \ P \triangleq P\sub{x_1,x_2}{\proj{M}, \projj{M}}
	\end{array}
	\]
	
	As an example of the introduced syntax, below we give the formal
	specification (that uses the equational theory $E_0$ from Fig.~\ref{fig_eqt}) for the roles in the BDH protocol presented in Fig.~\ref{fig_orig}.
	\begin{equation}
		\notag
		\begin{split}
			& \begin{aligned}
				\bdh{C}(&s, c, ch) \triangleq \nw a
				.\cout{ch}{\smult{a}{\pk{c}}}. \\
				& \cin{ch}{y}. \\
				& \lett \ \key{c} \ \coloneqq \hash{\smult{\mult{a}{c}}{y}} \texttt{in}\\
				& \mbox{\squeezespaces{1}$\lett \ 
					\certificate \ \coloneqq \pair{\pk{c}}{\sig{s}{\pk{c}}} \texttt{in}$} \\
				& \cout{ch}{\enc{\pair{\pair{a}{\pk{c}}}{
							\certificate}}{\key{c}}}
			\end{aligned}\\	
			\\
			& \begin{aligned}
				\bdh{T}(&pk_s, ch) \triangleq \nw t 
				.\cin{ch}{z_1}. \\
				& \cout{ch}{\pk{t}}. \\
				& \cin{ch}{z_2}. \\
				& \lett \ \key{t} \ \coloneqq \hash{\smult{t}{z_1}} \texttt{in} \\
				&  \lett \angles{m_1, m_2} \coloneqq \\ 
				& \angles{\proj{\dec{\key{t}}{z_2}}, \projj{\dec{\key{t}}{z_2}}} \texttt{in}\\
				& \mbox{\squeezespaces{0.9}$\texttt{if}\, \projj{m_1} = \checksig{pk_s}{\projj{m_2}} \texttt{then}$} \\ 
				& \texttt{if}\, \smult{\proj{m_1}}{\projj{m_1}} = z_1\,\texttt{then} \ \cout{ch}{\ack} \\
			\end{aligned}
		\end{split}
	\end{equation}
	
	The card role process is parametrised by the secret key $s$ of the payment system, the secret key $c$ of the card and the session channel $\mathit{ch}$. The terminal role is parametrised only by the system's public key $pk_s$ and $\mathit{ch}$. The action $\cout{ch}{\ack}$ is an event used to indicate at what point the terminal believes it has authenticated the card.
	
	\subsection{Semantics}
	
	We present the state of a process as an \emph{extended process} $\mathopen{\nu \vec{x}.}\left( \sigma
	\cpar P \right)$. The syntax for extended processes is given in
	Fig.~\ref{fig_extended}. An extended process comprises private values $\vec{x}$, playing the role of keys, nonces, fresh channels, etc., messages already sent on the network $\sigma$ and the \emph{future actions} $P$. For example, the extended process $\mathopen{\nu s.}\left( \sub{u_1, u_2}{\pks{s}, M} \cpar \cin{a}{z} \right)$ is composed of the fresh private secret key $s$, the sent messages $\pks{s}$ and $M$, and the input action $\cin{a}{z}$, that is not executed yet.
	\begin{figure}[h]
		\[
		\begin{array}{rlr}
			\mbox{Extended processes:} \\[3pt]
			A \Coloneqq& \sigma \cpar P & \\ 
			\mid \, & \mathopen{\nu x.} A & \\
		\end{array}
		\begin{array}{rlr}
			\mbox{Transition labels:} \\[3pt]
			\pi \Coloneqq & \tau  & \\
			\small| & \co{M}(z) & \\
			\small| & M\,N      & \\
		\end{array}
		\]
		\caption{A syntax for extended processes and transition labels.}
		\label{fig_extended}
	\end{figure}
	Notice that to list
	the messages sent we use the
	substitution $\sigma = \sub{u_1, \hdots, u_n}{M_1, \hdots, M_n}$,
	meaning that the message $M_i$ is available through the
	\quotes{alias} variable $u_i$. When a substitution serves as a ledger
	of sent messages, we refer to it as a \emph{frame}. We require
	extended processes $\mathopen{\nu \vec{x}.}\left(\sigma \cpar
	P\right)$ to be in normal form, i.e.\ to satisfy the restriction that the
	variables in $\dom{\sigma}$ are fresh for $\vec{x}$, $\fv{P}$ and
	$\fv{y\sigma}$, for all variables $y$. That is, $\sigma$ is
	idempotent, and substitutions are fully applied to $P$. We follow the
	convention that operational rules are defined directly on extended
	processes in normal form. This avoids numerous complications caused by
	the structural congruence in the original definition of bisimilarity
	for the applied $\pi$-calculus. 
	
	\begin{figure*}
		\[
		\begin{gathered}
			\begin{array}{l}
				\begin{prooftree}
					M\sigma \mathrel{=_{E}} K
					\justifies
					\sigma \cpar \mathopen{\cin{K}{y}.}P \lts{M\,N} \sigma \cpar {P\sub{y}{N\sigma}}
					\using
					\mbox{\textsf{Inp}}
				\end{prooftree}
				\qquad
				\begin{prooftree}
					\mbox{$u$ \fresh{} $M, N, P, \sigma$}
					\qquad
					M\sigma \mathrel{=_{E}} K
					\justifies
					\sigma \cpar \cout{K}{N}.P \lts{\co{M}(u)} {\sub{u}{N}}\circ\sigma \cpar P
					\using
					\mbox{\textsf{Out}}
				\end{prooftree}
				\qquad
				\begin{prooftree}
					\sigma \cpar P \lts{\pi} A 
					\qquad
					M \mathrel{=_{E}} N
					\justifies
					{\sigma \cpar \mathopen{\match{M = N}}{P}\lts{\pi}{A}}
					\using
					\mbox{\textsf{Mat}}
				\end{prooftree}
				\\[20pt]
				\begin{prooftree}
					A \lts{\pi} B
					\quad\enskip
					\mbox{$x$ \fresh{} $\n{\pi}$}
					\justifies
					{{\nu x.A}\lts{\pi}{\nu x.B}}
					\using
					\mbox{\textsf{Res}}
				\end{prooftree}
				\qquad
				\begin{prooftree}
					\sigma \cpar P \lts{\pi} B
					\quad\enskip
					\mbox{$x$ \fresh{} $\n{\pi}, \sigma$}
					\justifies
					\sigma \cpar \nu x.P \lts{\pi} {\nu x.B}
					\using
					\mbox{\textsf{Extrusion}}
				\end{prooftree}	
				\qquad
				\begin{prooftree}
					\sigma \cpar P \lts{\pi} \mathopen{\nu \vec{x}.}\left( \rho \cpar Q \right)
					\quad\enskip
					\mbox{$\vec{x} \cup \bn{\pi}$ \text{\fresh{}} $P$}
					\justifies
					\sigma \cpar \bang P \lts{\pi} \mathopen{\nu \vec{x}.}\left( \rho \cpar Q  \cpar \bang P \right)
					\using
					\mbox{\textsf{Rep-act}}
				\end{prooftree}
			\end{array}
			\\[10pt]
			\begin{array}{cc}
				\begin{prooftree}
					{\sigma \cpar P \lts{\pi} \mathopen{\nu \vec{x}.}\left( \rho \cpar R\right)}
					\qquad
					\mbox{$\vec{x} \cup \bn{\pi}$ \fresh{} $Q$}
					\justifies
					\sigma \cpar {P \cpar Q} \lts{\pi} \mathopen{\nu \vec{x}.}\left( \rho \cpar R \cpar Q\right)
					\using
					\mbox{\textsf{Par-l}}
				\end{prooftree}
				\qquad
				\begin{prooftree}
					{\sigma \cpar P \lts{\pi} \mathopen{\nu \vec{x}.}\left( \rho \cpar R\right)}
					\qquad
					\mbox{$\vec{x} \cup \bn{\pi}$ \fresh{} $Q$}
					\justifies
					\sigma \cpar {Q \cpar P} \lts{\pi} \mathopen{\nu \vec{x}.}\left( \rho \cpar Q \cpar R\right)
					\using
					\mbox{\textsf{Par-r}}
				\end{prooftree}
			\end{array}
		\end{gathered}
		\]
		\caption{A labelled transition system defined on extended processes in normal form.}
		\label{fig_transrules}
	\end{figure*}

	An extended process $\mathopen{\nu \vec{x}.}\left( \sigma \cpar P
	\right)$ may make a transition to a new state by executing an action
	available in $P$. We present transitions as labelled arrows. The
	syntax for labels is presented in Fig.~\ref{fig_extended}. To
	describe the transition rules we define the \emph{bound names} of the
	transition label such that $\bn{\pi} = \left\{ x \right\}$ only if
	$\pi = \co{M}(x)$ and $\bn{\pi} = \emptyset$ otherwise and the
	\emph{names} such that $\n{M\,N} = \fv{M} \cup \fv{N}$ and $\n{M(x)}
	= \fv{M} \cup \left\{x\right\}$. Finally, we present the transition
	rules in Fig.~\ref{fig_transrules}.
	The label of a transition
	represents an action that the process takes to arrive at a new state.
	In our reduced version of the applied $\pi$-calculus those actions are
	either input or output: specifically $M\,N$ denotes the input of
	message $N$ on channel $M$ and $\co{M}(z)$ denotes an output on
	channel $M$ of a message bound to the variable $z$. Notice the importance of capture avoidance in the rule \textsf{Inp}. For instance $\nw n, k. \left(\sub{u}{\enc{n}{k}} \cpar \cin{c}{x}.P\right)$ can execute an input action $c\,n$: since $n$ is a bound name, it is renamed using $\alpha$-conversion to e.g. $z$ in the initial process to avoid a name clash occurring when the substitution is applied to $P$ (bottom-right of the \textsf{Inp} rule). The resulting process is $\nw z, k \left(\sub{u}{\enc{z}{k}} \cpar P\sub{n, x}{z, n}  \right)$.
	
	Fig.~\ref{fig_transrules} is missing any rule for $\tau$
	transitions, invisible for external observers. We
	purposefully left these rules out since the protocols are modeled in a
	way that no $\tau$ actions can occur. The absence of $\tau$
	transitions, in turn, allows us to employ \emph{strong} semantics in
	the definition of quasi-open bisimilarity which simplifies the
	bisimilarity check since it ensures certain finiteness, i.e., each
	process has finitely many $\pi$-labelled transitions for any label
	$\pi$. Strong bisimilarity, in contrast to \emph{weak} bisimilarity,
	cares about the number of silent $\tau$ transitions which is always
	zero in this paper.
	
	\subsection{Equivalence notion}
	In Section~\ref{sec_unlink} we define the unlinkability of the payment system as an equivalence notion: if the system behaves like the ideal unlinkable system, then it is unlinkable. In this subsection we formally define the exact equivalence relation for extended processes that we use in the paper. 
	
	An equivalence captures both static and dynamic parts of processes' behaviour: no
	distinction is made for processes if they output the same sequence of
	messages so far and if they can match each other's actions. That is, we require such relation to be
	\emph{bisimilarity}. The exact type of bisimilarity is important though, and there are many notions of
	bisimilarity~\cite{sangiorgi01walker}. 
	We consider a bisimilarity that is also a \emph{congruence}.
	Recall that we study the unlinkability of the card in a hostile
	environment, hence we wish the unlinkability property to hold in
	\emph{any context}, e.g. in the absence or presence of any terminals.
	
	We start with a standard definition of static equivalence, that captures the distinction between two snapshots of the protocol execution and then will make our way to a bisimilarity congruence.
	
	\begin{definition}\emph{(static equivalence)}\label{def_static}
		Two extended processes $\mathopen{\nu\vec{x}.}\left( \sigma \cpar P \right)$ and $\mathopen{\nu\vec{y}.}\left( \theta \cpar Q \right)$ are statically equivalent whenever for all messages $M$ and $N$ such that $\vec{x}, \vec{y} \text{ \emph{\fresh{}} } M, N$, we have	$M\sigma =_{E} N\sigma$ if and only if $M\theta =_{E} N\theta$.
	\end{definition}
	
	In the context of the definition above, we say that the message $M$ is
	a \emph{recipe} for some message $K$ under $\sigma$ if $M \text{
		\fresh{} } \vec{x}$ and $M \sigma =_E K$.
	
	To ensure that our notion of bisimilarity is a congruence relation we require our bisimulation to be an open relation. A relation is \textit{open} if it is preserved under substitutions fresh for the domain of the frame of the extended process, as stated formally below. 
	By introducing this condition we give our attacker the capacity to
	influence messages bound to free variables (therefore accounting for
	all possible ways to ``stage'' the attack) without access to the outputs
	recorded in the frame -- they may only be used as a part of the
	input.
	
	\begin{definition}\emph{(open relation)}
		A relation $\mathrel{\mathcal{R}}$ over extended processes is \textit{open} whenever, if $A = \mathopen{\nu \vec{x}.}\left( \sigma \cpar P\right)$ and $B = \mathopen{\nu \vec{y}.}\left( \theta \cpar Q\right)$ and $A \mathrel{\mathcal{R}} B$, then for all $\rho$ such that $\dom{\sigma}$ is\footnote{\scriptsize$\dom{\sigma} = \dom{\theta}$ is an invariant property of a bisimulation, meaning that any pairs of processes not satisfying this property can be safely removed from a bisimulation.} fresh for $\rho$, we have $A\rho \mathrel{\mathcal{R}} B\rho$.
	\end{definition}
	
	The precise technical name for the notion of bisimilarity restricted to open relations is \emph{quasi-open bisimilarity} --- the coarsest of bisimilarities for the applied $\pi$-calculus that is a congruence~\cite{horne2021quasi}. We stress the importance of coarseness here: verifying against too fine equivalence may lead to spurious attacks.
	
	\begin{definition}\emph{(quasi-open bisimilarity)}\label{def_quasiopen}
		An open symmetric relation between extended processes $\mathrel{\mathcal{R}}$ is a quasi-open bisimulation whenever, if $A \mathrel{\mathcal{R}} B$ then the following hold: 
		\begin{itemize}
			\item $A$ and $B$ are statically equivalent.
			\item If $A \lts{\pi} A'$ there exists $B'$ such that $B \lts{\pi} B'$ and $A' \mathrel{\mathcal{R}} B'$.
		\end{itemize}
		Processes $P$ and $Q$ are quasi-open bisimilar, written $P \sim Q$,	whenever $P \mathrel{\mathcal{R}} Q$ for some quasi-open bisimulation $\mathcal{R}$.
	\end{definition}
	
	In the next section we define our unlinkability property by using
	quasi-open bisimilarity. The bisimilarity-based approach takes into
	account the fact that an attacker can make decisions during the
	execution of a protocol. Moreover, in comparison to familiar trace
	equivalence, for checking which tools like
	DeepSec~\cite{cheval2018rakotonirina} may help, bisimilarity is a
	\emph{safer} option since trace equivalence is coarser. Spelled out,
	this means that if a privacy property is defined using bisimilarity
	and it holds, then it holds when the bisimilarity is replaced by trace
	equivalence in the definition of the property. The opposite is not
	true as illustrated in related work using the BAC protocol for
	ePassport~\cite{filimonov2019breaking,horne2021mauw}, which is
	unlinkable if trace equivalence is instead employed, while, with
	respect to bisimilarity, there is a distinguishing strategy that can
	be exploited to link two sessions involving the same ePassport.
	Moreover, the openness/congruence feature of our chosen notion of bisimilarity allows us to prove properties in a modular way: an equivalence-based property that a smaller subsystem satisfies extends to a larger system for free.
	
	\subsection{Describing attacks as modal logic formulas}\label{sec_formulae}
	
	To conclude the background section we describe a succinct way of expressing attacks on bisimilarity. We will use a minimal fragment of a modal logic~\cite{Horne2018,horne2018coarse}, sufficient for the purpose of the paper. The syntax for formulae is very concise.
	\[
	\begin{array}{rlr}
		\psi \Coloneqq&  M = N & \mbox{equality} \\
		\mid& \diam{\pi}\psi & \hspace{21pt} \mbox{diamond} \\
	\end{array}
	\]
	The semantics of our minimal modal logic is as follows. 
	\[
	\begin{array}{lcl}
		\mathopen{\nu \vec{x}.}\left( \sigma \cpar P \right) \vDash M = N
		&\mbox{$\Leftrightarrow$}& M\mathclose{\sigma} \mathrel{=_E} N\mathclose{\sigma} \text{ and }  \vec{x} \ \text{\fresh{}} \ M, N
		\\
		A \vDash \diam{\pi}\psi &\mbox{$\Leftrightarrow$}& \exists B \text{, s.t. } A \lts{\pi} B \text{ and }B \vDash \psi
	\end{array}
	\]

	If there is a formula $\psi$ that is satisfied by one process, but is not satisfied by the other, e.g. $A \vDash \psi$, but $B \nvDash \psi$, then we know that $A \nsim B$ holds. The converse does not hold unless we take a larger modal logic, but this fragment suffices for the current protocol. The formula $\psi$ captures the strategy of an attacker for distinguishing two processes. Such a distinguishing strategy is a trace of transitions that the process $A$ can make, but the process $B$ may fail to match, followed by a test $M=N$ demonstrating the violation of static equivalence.
	We will use this modal logic approach to formally present our previously mentioned attack on the BDH protocol in the proof of Theorem~\ref{thm_link} in the next section. 
	
	\section{Unlinkability}\label{sec_unlink} 
	In this section we introduce a formal definition of unlinkability as a process equivalence, and show that the BDH protocol in Fig.~\ref{fig_orig} does not satisfy this definition.

	\subsection{Verification of unlinkability is challenging}
	
	There is a certain feature in the original definition of (strong) unlinkability proposed by Arapinis et al.~\cite{arapinis2010analysing}, namely the use of \emph{weak} transitions in the underlying bisimilarity notion. In the weak semantics for a given process $A$ and the transition label $\pi$ there could be infinitely many states $B$ s.t. $A \lts{\pi} B$ which can make verification a difficult task. 
	
	To overcome this obstacle we follow a method developed by Horne and
	Mauw~\cite{horne2021mauw} allowing the reduction of weak to strong
	bisimilarity that supports image finiteness. The insight of their work
	is that a certain way of expressing a protocol in the applied $\pi$-calculus makes verification easier without compromising unlinkability in the original sense. We adopt this method not only out of safety consideration: bisimilarity is stronger than trace equivalence and we are not losing anything when verifying in a stronger setting, but also to open up a discussion on possible automation of checking the bisimilarity of two processes. Studying the detailed proof of bisimilarity we provide in Section~\ref{sec_theorem} may outline the steps and challenges to overcome when considering tool support.
	
	Another, parallel, approach to 
	unlinkability verification is rolling back to a familiar trace equivalence, an equivalence with established tool support, by proving that a particular class of protocols indeed allows doing that. For instance, this approach was taken recently by Baelde, Delaune, and Moreau to analyse stateful protocols~\cite{baelde2020unlink}. This work also demonstrates limitations of our method: analysing BDH, we \emph{can} drop one party, the terminal, from consideration because cards and terminals share no secret, while otherwise, an observed reaction of an honest participant may break unlinkability.

	\subsection{Definition of unlinkability}
	
	Perhaps the most straightforward way to design unlinkable payments is to introduce cards that immediately expire after one use. Such cards can never participate in a purchase more than once and payments are undoubtedly unlinkable. We say that if the \quotes{real world} system where cards are used multiple times, is indistinguishable by an attacker from an idealised unlinkable world in which cards are disposed of after each use, then unlinkability of payments is achieved.
	
	Let $\card(s, c, ch)$ be the card process scheme parametrised by the payment system's secret key $s$, communication channel $ch$ and the card's secret key $c$. 
	Then we have the following.
	
	\begin{definition}\emph{(unlinkability)} A card process scheme $C$ is unlinkable whenever
		\[
		\begin{array}{cc}
			\nw s.\cout{out}{\pks{s}}.\bang\nw c.\nw
			ch.\cout{card}{ch}.\card(s, c, ch) \\[0.5ex]
			\sim \\[0.5ex]
			\nw s.\cout{out}{\pks{s}}.\bang\nw c.\bang\nw ch.\cout{card}{ch}.\card(s, c, ch)
		\end{array}
		\]
		\label{def_unlink}
	\end{definition}
	The process on the left of the above relation models the idealised world where a card participates in no more than one transaction. This process starts by creating the secret key of the payment system $s$. Then the public key $\pks{s}$ of the payment system is made available via the output on the public channel $\mathit{out}$. Each newly manufactured card $c$ is allowed to participate in the execution of the payment protocol just once. The process on the right of the above relation models the more realistic situation where each card $c$ may participate in several runs of the protocol. If the idealised situation is equivalent to the real world one, where the equivalence we employ is quasi-open bisimilarity (Def.~\ref{def_quasiopen}), we say that the payment system satisfies unlinkability. 
	Since our chosen notion of equivalence is a congruence, we can
	check unlinkability for a subsystem comprising cards only and be sure
	that the presence of any terminals would not make the whole system
	linkable. This can be illustrated by the context
	\[
	\begin{array}[t]{l}
		\mbox{$\squeezespaces{0.35}\nu \, out.\left(\left\{ \cdot \right\}
			\cpar 
			\cin{out}{pk_s}. \cout{out'}{pk_s}\bang \nu ch_t. \cout{\tch}{ch_t}.\terminal(pk_s, ch_t) \right)$}
	\end{array}
	\]
	where $out'$ is a public channel used in a full system to
	announce the system's public key. The detailed proof that this
	context indeed leads to a correct representation of a full system with
	cards and terminals is given in the companion
	paper~\cite{horne2021quasi} dedicated entirely to the notion of
	quasi-open bisimilarity and containing proofs that
	Def.~\ref{def_quasiopen} is correct in the sense that it is a sound and complete congruence with respect to an established testing semantics.
	The contributions of the current paper and the above mentioned companion paper are disjoint, since the paper you are now reading focusses on using quasi-open bisimilarity to verify a protocol.
	
	In summary, the process scheme on the left of the equation in Def.~\ref{def_unlink} is the specification and on the right is the implementation. If we can prove that the equivalence problem holds for a particular protocol then that protocol complies with the specification.
	This is similar to the pattern for specifying unlinkability introduced in related work~\cite{arapinis2010analysing}, with the key difference being that only cards need be accounted for, since the only information shared with the terminal is the public key of the payment system.
	
	In both cases in the definition above the card uses a newly created session channel $ch$ that is output on the public channel $\mathit{card}$, hence an attacker has the power to observe and control the creation of radio frequency communication channels. Moreover, a dedicated channel per session is a requirement in the transport protocol ISO/IEC 14443~\cite{iso14443} that is used in contactless EMV cards. Conveniently, session channels are the reason behind the lack of silent $\tau$-transitions in the protocols we study in the paper, which allows us to employ the strong notion of bisimilarity and simplifies proofs.
	
	\subsection{BDH is not unlinkable}
	Given the formal definition of unlinkability in Def.~\ref{def_unlink} we can now establish that the Blinded Diffie-Hellman protocol from Fig.~\ref{fig_orig} is not unlinkable.
	
	\begin{theorem} \begin{upshape}$\bdh{C}(s, c, ch)$\end{upshape} violates unlinkability.  \label{thm_link}
		\begin{proof}
			To describe the attack on unlinkability of the BDH protocol we follow the modal logic formula notation described in Section~\ref{sec_formulae}.
			Consider the following processes, 
			where $\bdh{\card}$ is as defined in Section~\ref{sec_syntax}.
			\[
			\begin{array}{l}
				\spec{RFC} \triangleeq \nw s.\cout{out}{\pks{s}}.\bang\nw c.\nw ch.\cout{card}{ch}.\bdh{\card}
				\\[4pt]
				\impl{RFC} \triangleeq \nw s.\cout{out}{\pks{s}}.\bang\nw c.\bang\nw ch.\cout{card}{ch}.\bdh{\card}
			\end{array}
			\]
			To show that $\spec{RFC} \nsim \impl{RFC}$ we present a formula that is satisfied by $\impl{RFC}$, but not by $\spec{RFC}$. Let the formula $\psi$ be as follows.
			\[
			\begin{array}{l}
				\diam{\co{out}(pk_s)}
				\\
				\diam{\co{card}(u_1)}\diam{\co{u_1}(v_1)}\diam{u_1\,\smult{y_1}{\gen}}\diam{\co{u_1}(w_1)}
				\\
				\diam{\co{card}(u_2)}\diam{\co{u_2}(v_2)}\diam{u_2\,\smult{y_2}{\gen}}\diam{\co{u_2}(w_2)}
				\\
				\big( \projj{\dec{\hash{\smult{y_1}{v_1}}}{w_1}} = \\
				\qquad\qquad\qquad\qquad\qquad\quad\enspace \, \projj{\dec{\hash{\smult{y_2}{v_2}}}{w_2}} \big)
			\end{array}
			\]
			
			The above formula describes two sessions of the BDH protocol, which, for $\impl{RFC}$, can be with the same card, say $c_1$. The equality test at the end of $\psi$ compares the certificates obtained from each session to each other, which the terminal can decrypt in both sessions. This certificate can be the same for both sessions of $\impl{RFC}$ involving the same card, since it is bound to the card's identity $c_1$. Therefore $\impl{RFC} \vDash \psi$. In contrast, $\spec{RFC} \nvDash \psi$ since every session is with a new card and hence the equality test never holds, since the certificates will always differ.
		\end{proof}
	\end{theorem}
	In the next section, we present our enhanced BDH protocol that satisfies unlinkability.
	
	\section{Making Blinded Diffie-Hellman truly unlinkable}\label{sec_making}
	In this section, we propose our improvement to the BDH protocol proposed by EMVCo  called \emph{Unlinkable BDH} (UBDH). This improvement makes use of a certification scheme with certificates invariant under blinding. We point to an existing instance of such a certification scheme, the Verheul certification scheme, and, finally, we prove that our improvement indeed makes the Blinded Diffie-Hellman protocol unlinkable~\cite{rfc}. 
	
	\subsection{Blinded Diffie-Hellman with blinded certificates}\label{sec_fixed}
	Recall from Section~\ref{sec_notunlinkable} and Theorem~\ref{thm_link} that
	the reason behind the failure of unlinkability
	of the BDH protocol proposed by EMVCo is that the card gives away its static certificate and its blinding factor. While this allows an honest terminal to authenticate the card, the public key of the card ultimately obtained by the terminal may be used to track the card in the future. We demonstrate in this section that authentication can still be performed without disclosing the public key or the signature. In order to achieve this, we specify more precisely the signature scheme (initially unspecified by EMVCo) used for certificate verification. In particular, we require that blinding and signing operations must commute. In this case, the signature can be blinded with the same nonce as the card's public key at the beginning of the session and later checked against the public key of the payment system directly in its blinded form. As a result, only the blinded version of the card's public key is ever revealed.
	
	The equational theory $E$ for the improved protocol is the equational
	theory $E_0$ in Fig.~\ref{fig_eqt} extended with the
	property expressed in Fig.~\ref{fig_blinding}, which permits scalar multiplication and signing to commute.
	\begin{figure}[h]
		\[
		\smult{M}{\sig{K}{N}} =_E \sig{K}{\smult{M}{N}}
		\]
		\caption{Equation for blinding extending the equational theory in
			Fig.~\ref{fig_eqt}.}\label{fig_blinding}
	\end{figure}
	
	It now follows from the blinding condition above and the last equation in Fig.~\ref{fig_eqt} that the check of the signature, blinded with some blinding factor, returns the message, blinded with the same factor. This property of signatures in the equational theory $E$ is used by the terminal when authenticating the card in our proposed update of the BDH protocol. The updated BDH protocol is presented informally in Fig.~\ref{fig_impr} and the corresponding formal $\pi$-calculus specification of the two roles involved is presented below.
	\begin{figure}[t]
		\includegraphics[width=\linewidth]{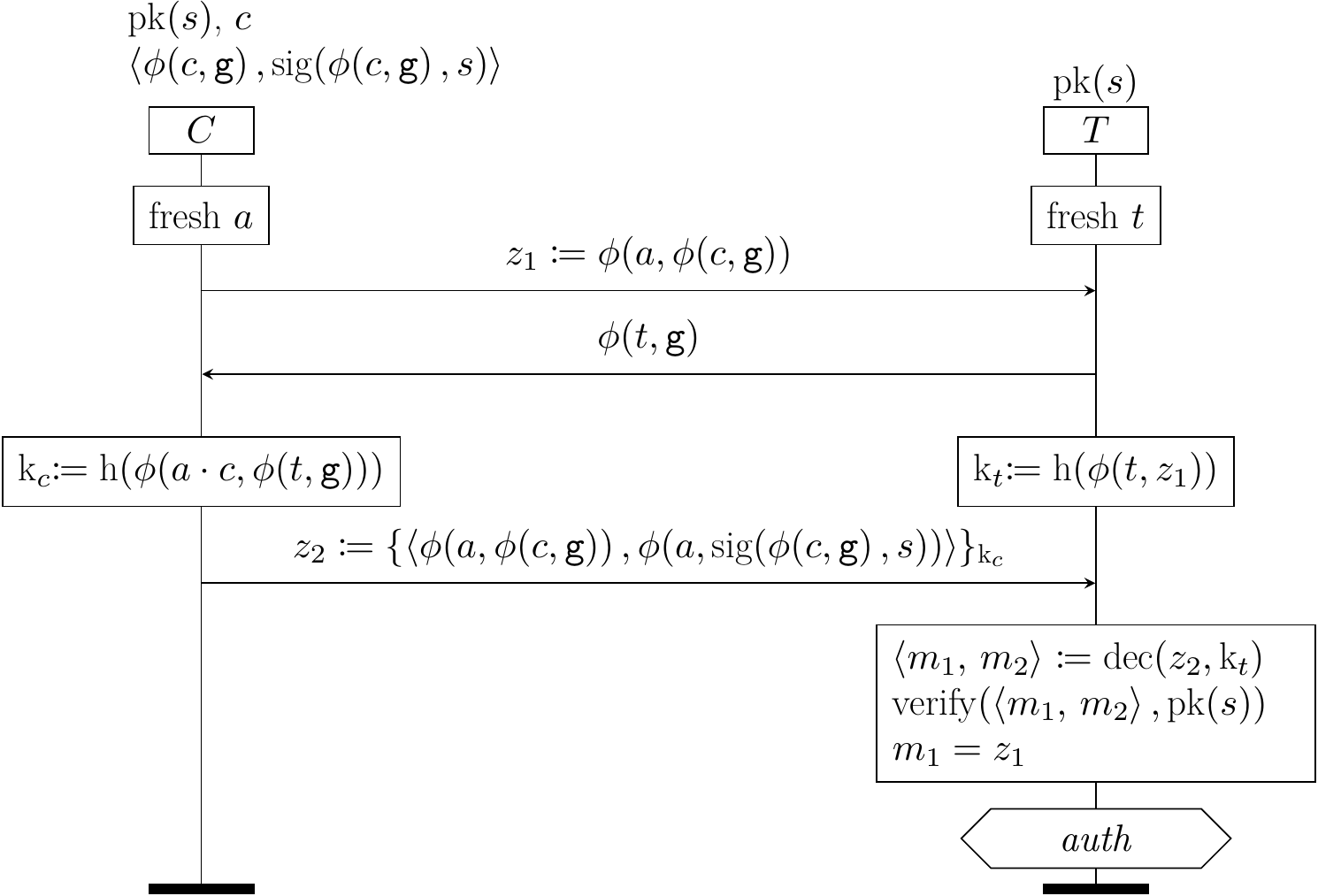}
		\centering
		\caption{The Unlinkable BDH protocol.} 
		\label{fig_impr}
	\end{figure}
	\begin{equation}\label{eq_fixed}
		\notag
		\begin{split}
			& \begin{aligned}
				\fix{C}(&s, c, ch) \triangleq \nw a.\cout{ch}{\smult{a}{\pk{c}}}.\\
				& \cin{ch}{y}.\\
				& \lett \ \key{c} \coloneqq \hash{\smult{\mult{a}{c}}{y}} \texttt{in} \\
				& \mbox{$\squeezespaces{0.7}\lett \ m \coloneqq \pair{\smult{a}{\pk{c}}}{\smult{a}{\sig{s}{\pk{c}}}} \texttt{in}$} \\
				& \cout{ch}{\enc{m}{\key{c}}}
			\end{aligned}\\	
			\\[0.01ex]
			& \begin{aligned}
				\fix{T}(&pk_s, ch) \triangleq \nw t.\cin{ch}{z_1}. \\
				& \cout{ch}{\pk{t}}. \\
				& \cin{ch}{z_2}. \\
				& \lett \ \key{t} \coloneqq \hash{\smult{t}{z_1}} \texttt{in}\\
				& \lett \angles{m_1, m_2} \coloneqq \\ 
				& \angles{\proj{\dec{\key{t}}{z_2}}, \projj{\dec{\key{t}}{z_2}}}\, \texttt{in} \\ 
				& \texttt{if}\, m_1 = \checksig{pk_s}{m_2}\, \texttt{then} \\
				& \texttt{if}\, m_1 = z_1\, \texttt{then} \ \cout{ch}{\ack}  \\
			\end{aligned}
		\end{split}
	\end{equation}
	
	Our version differs from the original proposal in message $z_2$
	sent by the card to the terminal, i.e. now only the (encrypted)
	blinded certificate is transferred. 
	At no point in the protocol, can the terminal
	unblind the card's public key since the blinding factor $a$ is never
	revealed to any terminal.

	We conclude this subsection by mentioning that there is a signature scheme satisfying both the blinding condition in Fig.~\ref{fig_blinding}, and the technical requirements of the BDH protocol~\cite{rfc}, namely the Verheul certification scheme~\cite{verheul2001self}. The scheme has been implemented on smart cards~\cite{batina2010developing} by Batina et al., using BN\footnote{\scriptsize The original paper~\cite{verheul2001self}, in contrast, describes the system using symmetric pairings on a supersingular curve. This approach
		historically precedes the asymmetric one, making certain Decisional
		Diffie-Hellman problem simple and requires greater field size
		(which would slow down on-card computation) to achieve the same level
		of security as a non-supersingular curve based
		system~\cite{freeman2010scott}.} curves~\cite{barreto2006naehrig}
	with time presenting one blinded certificate of 0.45 seconds, which is within the limit of 500ms of the card present in the reader field~\cite{emvcontactless}. In the proposal~\cite{rfc} EMVCo intends to use p256 curve,
	however switching over to a pairing-friendly BN curve would not introduce any slow-downs, compared to p256 curve, in on-card computation as was shown by Dzurenda et al. in the performance analysis~\cite{dzurenda2017ricci} of different elliptic curves on smart cards.

	\subsection{Self-blindable certificates bring unlinkability in BDH}\label{sec_theorem}
	In this section we present a detailed proof of unlinkability of the
	Unlinkable BDH protocol that will illustrate the
	importance of the chosen equivalence relation (quasi-open bisimilarity, Def.~\ref{def_quasiopen}). We define $\spec{UPD}$ and $\impl{UPD}$ as
	\[
	\begin{array}{l}
		\spec{UPD} \triangleeq \nw s.\cout{out}{\pks{s}}.\bang\nw c.\nw ch.\cout{card}{ch}.\fix{\card}(s, c, ch)
		\\[4pt]
		\impl{UPD} \triangleeq \nw s.\cout{out}{\pks{s}}.\bang\nw c.\bang\nw ch.\cout{card}{ch}.\fix{\card}(s, c, ch)
	\end{array}
	\]
	
	The UBDH protocol is unlinkable as established by the following theorem. 
	\begin{theorem}\label{thm_static}
		\begin{upshape}$\fix{C}(s, c, ch)$\end{upshape} satisfies unlinkability. 	
		\begin{proof}		
			By Def.~\ref{def_unlink} of unlinkability, we must show that $\spec{UPD} \sim \impl{UPD}$. Therefore we shall provide a quasi-open bisimulation relation $\mathfrak{R}$ such that $\spec{UPD} \ \mathfrak{R} \ \impl{UPD}$. 
			
			To define such $\mathfrak{R}$ we have to introduce some notation. Let
			$L, \, D \in \mathbb{N}$ be the number of sessions and the number of
			cards in the system, respectively. We use indices $l \in \{1, \hdots,
			L\}$ and $d \in \{1, \hdots, D\}$ to track sessions and cards. 
			
			Define $m^d(a, y)$ as the encrypted blinded certificate:
			\begin{equation}\notag
				\mbox{$\squeezespaces{0.00001}m^d(a, y) \coloneqq \enc{\pair{\smult{a}{\pk{c_d}}}{\smult{a}{\sig{s}{\pk{c_d}}}}}{\hash{\smult{\mult{a}{c_d}}{y}}}$} 
			\end{equation}
			
			Define a partition $\Psi \coloneqq \{\alpha, \, \beta, \, \gamma\, \,
			\delta\}$ of the set of all sessions $\{1, \hdots, L\}$, where
			$\alpha$ is the set of sessions in which the channel is created, but
			no message has been sent; $\beta$ is the set of sessions in which the
			blinded public key has been sent but the response has not been
			received; $\gamma$ is the set of all sessions in which the response
			has been received but the encrypted blinded certificate has not been sent; $\delta$ is the set of all sessions in which the encrypted blinded certificate has been sent.
			
			Define a partition $\Omega \coloneqq \{\zeta^1, \hdots, \zeta^D \}$
			of the set of all sessions $\{1, \hdots, L\}$, where $\zeta^d$ is the set of all sessions with the card $d$. 
			
			Let $\vec{Y} \coloneqq (Y_1, \hdots, Y_L)$ be the list of inputs,
			where $Y_l$ is the input in session $l$. Recall that $Y_l$ can
			refer to messages already output on the network (the last line in
			Fig.~\ref{fig_relation}). Let $K \coloneqq \lvert \beta \cup \gamma \cup \delta \rvert$ be the number of \emph{started} sessions. Since we consider processes up to $\alpha$-conversion and permutation of names (aka.~equivariance), we assume that $a_l$ is the blinding factor in session $l$. 
			
			Finally, we define the following process subterms, which correspond to the elements of the partition $\Psi$.
			\begin{equation}
				\notag
				\begin{split}
					& \begin{aligned}
						\m{E}^d(ch) \triangleq \nw a. \cout{ch}{\smult{a}{\pk{c_d}}}.\m{F}^d(ch, a)
					\end{aligned}\\
					& \begin{aligned}
						\m{F}^d(ch, a) \triangleq \cin{ch}{y}. \m{G}^d(ch, a, y)
					\end{aligned}\\
					& \begin{aligned}
						\m{G}^d(ch, a, y) \triangleq \mbox{$\squeezespaces{1}\cout{ch}{m^d(a, y)}$}
					\end{aligned}\\
					& \begin{aligned}
						\m{H}^d \triangleq 0
					\end{aligned}
				\end{split}
			\end{equation}
			
			The bisimulation relation $\mathfrak{R}$ is defined as \emph{the
				least symmetric open relation} satisfying the
			constraints\footnote{\scriptsize The relation $\mathfrak{R}$ may not be the
				\emph{smallest} quasi-open bisimilarity satisfying $\spec{UPD} \
				\mathfrak{R} \ \impl{UPD}$.}
			in Fig.~\ref{fig_relation}. Spelled out, we pair the reachable states of $\spec{UPD}$ and $\impl{UPD}$ based on the number of sessions and the respective stages of the card in a session. Notice that $\spec{UPD} \ \mathfrak{R} \ \impl{UPD}$ by the definition of $\mathfrak{R}$.
			
			\begin{figure}[h]
				\[
				\begin{array}{c}
					\spec{UPD} \ \mathfrak{R} \ \impl{UPD} \\[4pt]
					\begin{array}{l}
						\spec{UPD}^{\Psi}(\vec{Y}) \triangleeq \nu s, c_1, \hdots, c_L, ch_1, \hdots, ch_L, \\[2pt]
						a_{l_1}, \hdots, a_{l_K}.(\sigma
						\\[4pt]
						\begin{array}[t]{l}
							\cpar C_1 \cpar \hdots \cpar C_L \\[4pt]
							\cpar \bang \nw c.\nw ch.\cout{card}{ch}.\fix{\card}(s, c, ch))
						\end{array}
					\end{array} \\[30pt]
					\begin{array}{c}
						\mathfrak{R}
					\end{array} \\[1pt]
					\begin{array}{l}
						\impl{UPD}^{\Psi, \Omega}(\vec{Y}) \triangleeq \nu s, c_1, \hdots, c_D, ch_1, \hdots, ch_L, \\[2pt]
						a_{l_1}, \hdots, a_{l_K}.(\theta
						\\[4pt]
						\begin{array}[t]{l}
							\mbox{$\squeezespaces{1}\cpar \hdots \cpar C_{l}^d \cpar \hdots \cpar \bang \nw ch.\cout{card}{ch}.\fix{C}(s, c_d, ch)$}
							\\[6pt]
							\cpar \bang \nw c.\bang\nw ch.\cout{card}{ch}.\fix{\card}((s, ch, c)))
						\end{array}
					\end{array}
				\end{array}
				\]
				\[
				\arraycolsep=2pt
				\begin{array}{l}
					C_l = \left\{
					\begin{array}{ll}
						\m{E}^l(ch_l) & \mbox{ if $l \in \alpha$} 
						\\[2pt]
						\m{F}^l(ch_l, a_{l}) & \mbox{ if $l \in \beta$} 
						\\[2pt]
						\m{G}^l(ch_l, a_{l}, Y_{l}\sigma) & \mbox{ if $l \in \gamma$} 
						\\[2pt]
						\m{H}^l & \mbox{ if $l \in \delta$} 
					\end{array}
					\right.
					\\[9pt]
					C_{l}^d = 
					\left\{
					\begin{array}{ll}
						\m{E}^d(ch_l) & \mbox{ if $l \in \zeta^d \cap \alpha$} 
						\\[2pt]
						\m{F}^d(ch_l, a_{l}) & \mbox{ if $l \in \zeta^d \cap \beta$} 
						\\[2pt]
						\m{G}^d(ch_l, a_{l}, Y_{l} \theta) & \mbox{ if $l \in \zeta^d \cap \gamma$} 
						\\[2pt]
						\m{H}^d & \mbox{ if $l \in \zeta^d \cap \delta$} 
					\end{array}
					\right.
					\\
					\begin{array}{ll}
						pk_s \sigma = \pks{s}
						\\
						u_l \sigma = ch_l & \mbox{ if $l \in \{1, \hdots, L\}$}
						\\
						v_l \sigma = \smult{a_l}{\pk{c_l}} & \mbox{ if $l \in \beta \cup \gamma \cup \delta$}
						\\
						w_l \sigma = m^l(a_l, Y_l \sigma) & \mbox{ if $l \in \delta$}
						\\[8pt]
						pk_s \theta = \pks{s}
						\\
						u_l \theta = ch_l & \mbox{ if $l \in \{1, \hdots, L\}$}
						\\
						v_l \theta = \smult{a_l}{\pk{c_d}} & \mbox{ if $l \in \zeta^d \cap \left(\beta \cup \gamma \cup \delta \right)$}
						\\
						w_l \theta = m^d(a_l, Y_l  \theta) & \mbox{ if $l \in \zeta^d \cap \delta$}
					\end{array}
				\end{array}
				\]
				\[	\mbox{$\squeezespaces{0.5}\Psi \coloneqq \{\alpha, \, \beta, \, \gamma, \, \delta\}$}, \ \, \mbox{$\squeezespaces{0.5}\Omega \coloneqq \{\zeta^1, \hdots, \zeta^D \}$} \text{ are partitions of $\squeezespaces{0.1}\{1, \hdots, L\}$}\]
				\[K \coloneqq \lvert \beta \cup \gamma \cup \delta \rvert\ \quad  l_1,\hdots,l_K \in {\beta \cup \gamma \cup \delta}\]
				\[pk_s, u_l, v_l, w_l \text{ \fresh{} } \{card, s\} \cup \{c_l, ch_l, a_l | l \in \{1, \hdots, L\}\}\]
				\[Y_l \text{ \fresh{} } \{s\} \cup \{c_l, ch_l, a_l | l \in \{1, \hdots, L\}\}\]
				\[\fv{Y_l} \cap \left( \{v_i | i \in \alpha\} \cup \{w_i | i \in \alpha \cup \beta \cup \gamma \cup \{l\}\}\right) = \emptyset\]
				\caption{Defining conditions for the bisimulation relation $\mathfrak{R}$. }
				\label{fig_relation}
			\end{figure}
			
			To prove that $\mathfrak{R}$ is indeed a quasi-open bisimulation, according to Def.~\ref{def_quasiopen}, we must demonstrate
			\begin{enumerate}
				\item \emph{(bisimulation)} Whenever $A \ \mathfrak{R} \ B$, and $A \lts{\pi} A'$, there exists $B'$ such that $B \lts{\pi} B'$ and $A' \ \mathfrak{R} \ B'$.
				\item \emph{(openness)} $\mathfrak{R}$ is closed under the application 
				of a substitution fresh for the domain of the frame of any of the related states.
				\item \emph{(static equivalence)} Whenever $A \ \mathfrak{R} \ B$, $A$ is statically equivalent to $B$.
			\end{enumerate}
			
			Since $\mathfrak{R}$ is by definition a symmetric relation, we provide proof only for the cases when the left-side process starts first. Below we present the exhaustive list of cases for the defining conditions of the relation $\mathfrak{R}$ in Fig.~\ref{fig_relation}. Proof trees justifying each transition can be found in Appendix~\ref{app_prooftrees}. Openness and static equivalence are discussed separately.
			
			\textit{Case} 1. $\spec{UPD} \ \mathfrak{R} \ \impl{UPD}$, $\co{out}(pk_s)$. The process $\spec{UPD}$ can do the transition $\co{out}(pk_s)$ to the state $\spec{UPD}^{\emptyset}(\emptyset)$. There is a state $\impl{UPD}^{\emptyset, \emptyset}(\emptyset)$ to which the process $\impl{UPD}$ can do the transition $\co{out}(pk_s)$. By the definition of $\mathfrak R$ we have $\spec{UPD}^{\emptyset}(\emptyset)\ \mathfrak{R} \ \impl{UPD}^{\emptyset,\emptyset}(\emptyset)$.
			
			\textit{Case} 2. $\spec{UPD}^{\Psi}(\vec{Y}) \ \mathfrak{R} \ \impl{UPD}^{\Psi, \Omega}(\vec{Y})$, $\co{card}(u_{L+1})$. The process $\spec{UPD}^{\Psi}(\vec{Y})$ can do the transition $\co{card}(u_{L+1})$ to the state $\spec{CH} \triangleeq \spec{UPD}^{\{\alpha \cup \{L+1\}, \beta, \gamma, \delta\}}((Y_1, \hdots, Y_L, \emptyset))$. In the process $\impl{UPD}^{\Psi, \Omega}$ either some card $d$ starts a new session and the resulting state is $\impl{CH} \triangleeq \impl{UPD}^{\{\alpha \cup \{L+1\}, \beta, \gamma, \delta\}, \{\hdots, \zeta^d \cup \{L+1\} , \hdots\}}((Y_1, \hdots, Y_L, \emptyset))$ or the new card is created and the resulting state is $\impl{CHC} \triangleeq \impl{UPD}^{{\{\alpha \cup \{L+1\}, \beta, \gamma, \delta\}}, \Omega \cup \{\{L+1\}\}}((Y_1, \hdots, Y_L, \emptyset))$. In both cases by the definition of $\mathfrak{R}$ we have $\spec{CH} \ \mathfrak{R} \ \impl{CH}$ 
			and $\spec{CH} \ \mathfrak{R} \ \impl{CHC}$.
			
			\textit{Case} 3. $\spec{UPD}^{\Psi}(\vec{Y}) \ \mathfrak{R} \ \impl{UPD}^{\Psi, \Omega}(\vec{Y})$, $\co{u_l}(v_l)$, and $l \in \alpha$. The process $\spec{UPD}^{\Psi}(\vec{Y})$ can do the transition $\co{u_l}(v_l)$ to the state $\spec{APK} \triangleeq \spec{UPD}^{\{\alpha \setminus\{l\}, \beta \cup \{l\}, \gamma, \delta\}}(\vec{Y})$. There is a state $\impl{APK} \triangleeq \impl{UPD}^{\alpha \setminus \{l\}, \beta \cup \{l\}, \gamma, \delta\}, \Omega}(\vec{Y})$ to which the process $\impl{UPD}^{\Psi, \Omega}(\vec{Y})$ can do the transition $\co{u_l}(v_l)$. By the definition of $\mathfrak{R}$ we have $\spec{APK} \ \mathfrak{R} \ \impl{APK}$.
			
			\textit{Case} 4. $\spec{UPD}^{\Psi}(\vec{Y}) \ \mathfrak{R} \ \impl{UPD}^{\Psi, \Omega}(\vec{Y})$, $u_l \, Y_l$, and $l \in \beta$. 
			Let $\chi_l(\vec{Y}, M)$ be the list of message terms obtained from $\vec{Y}$ by the replacement of $l$th entry in $\vec{Y}$ with $M$. The process $\spec{UPD}^{\Psi}(\vec{Y})$ can do the transition $u_l \, Y_l$ to the state $\spec{IN} \triangleeq \spec{UPD}^{\{\alpha, \beta \setminus \{l\}, \gamma \cup \{l\}, \delta\}}(\chi_l(\vec{Y}, Y_l))$. There is a state $\impl{IN} \triangleeq \impl{UPD}^{\{\alpha, \beta \setminus \{l\}, \gamma \cup \{l\}, \delta\}, \Omega}(\chi_l(\vec{Y}, Y_l))$ to which the process $\impl{UPD}^{\Psi, \Omega}(\vec{Y})$ can do the transition $u_l \, Y_l$. By the definition of $\mathfrak{R}$ we have $\spec{IN} \ \mathfrak{R} \ \impl{IN}$.
			
			\textit{Case} 5. $\spec{UPD}^{\Psi}(\vec{Y}) \ \mathfrak{R} \ \impl{UPD}^{\Psi, \Omega}(\vec{Y})$, $\co{u_l}(w_l)$, and $l \in \beta$. The process $\spec{UPD}^{\Psi}(\vec{Y})$ can do a transition $\co{u_l}(w_l)$ to the state $\spec{CRT} \triangleeq \spec{UPD}^{\{\alpha, \beta, \gamma \setminus \{l\}, \delta \cup \{l\}\}}(\vec{Y})$. There is a state $\impl{CRT} \triangleeq \impl{UPD}^{\{\alpha, \beta, \gamma \setminus \{l\}, \delta \cup \{l\}\}, \Omega}(\vec{Y})$  to which the process $\impl{UPD}^{\Psi, \Omega}(\vec{Y})$ can do a transition $\co{ch_l}(w_l)$. By the definition of $\mathfrak{R}$ we have $\spec{CRT} \ \mathfrak{R} \ \impl{CRT}$.
			
			\textit{Openness}. $\mathfrak{R}$, by definition, is open: whenever $A \ \mathfrak{R} \ B$, then $A\rho \ \mathfrak{R} \ B\rho$ for any $\rho$ fresh for the domain of the frame of $A$. No such substitution $\rho$ introduce transitions not considered above. Indeed, since $\fv{\spec{UPD}} = \fv{\impl{UPD}} = \{out, card\}$, the substitution $\rho$ may only affect $out$, $card$ and free variables in the input $Y_l$. Therefore it is straightforward to modify proof trees [see the repository~\cite{repo} for details]: the transition label $\co{out}(pk_s)$ is replaced by $\co{out\rho}(pk_s)$, the transition label $\co{card}(u_{L+1})$ with $\co{card\rho}(u_{L+1})$ and $\vec{Y}$ with $\vec{Y}\rho \coloneqq \rounds{Y_1\rho, \hdots, Y_L\rho}$, where $\emptyset\rho \coloneqq \emptyset$. By $\alpha$-conversion we may assume that the range of $\rho$ does not contain variables \quotes{reserved} for future outputs or private nonces: $\fv{y\rho} \cap \left(\{s, pk_s\} \cup \{c_i, ch_i, a_i, u_i, v_i, w_i | l \in \mathbb{N}\}\right) = \emptyset$ for any $y$. Then, freshness conditions remain untouched up to the renaming of variables directly affected by $\rho$.
			
			\textit{Static equivalence}. To conclude, we prove that $A$ is statically equivalent to $B$ whenever $A \ \mathfrak{R} \ B$. There is nothing to prove in the case of $\spec{UPD} \ \mathfrak{R} \ \impl{UPD}$ since frames 
			are empty. The proof for the case $\spec{UPD}^{\Psi}(\vec{Y}) \ \mathfrak{R} \ \impl{UPD}^{\Psi, \Omega}(\vec{Y})$ is presented separately in Lemma~\ref{lemma_static}. 
		\end{proof}
	\end{theorem}
	
	To prove that $\spec{UPD}^{\Psi}(\vec{Y})$ is statically equivalent to $\impl{UPD}^{\Psi, \Omega}(\vec{Y})$ we use a weak notion of the \emph{normal form} $\nf{M}$ of a message term $M$, that captures the least complex, up to multiplication, expression of $M$; and the notion of the \emph{normalisation} of a frame $\sigma$ with respect to the equational theory $E$, which is a saturation of the range of $\sigma$ with weak normal forms of messages that have a recipe under $\sigma$. Recipes can be conveniently recorded in the domain of normalisation. Definitions are standard and given in Appendix~\ref{app_norm}.
	
	\begin{definition}($m$-atomic, $\phi$-atomic)
		A message term $M$ is $m$-atomic if there are no such $M_1$,
		$M_2$, s.t. $M =_E \mult{M_1}{M_2}$; it is $\phi$-atomic if there are
		no such $M_1$, $M_2$, s.t. $M =_E \smult{M_1}{M_2}$
	\end{definition}
	A subterm $N$ of $M$ is an \emph{immediate $m$-factor} if it is $m$-atomic and there is a message term $K$, s.t. $\mult{N}{K} = M$. 
	\begin{definition}(non-trivial recipe)
		The recipe $M$ is \emph{non-trivial} under $\sigma$ if $\fv{\nf{M}} \cap \dom{\sigma} \neq \emptyset$.
	\end{definition}
	We conclude the proof of Theorem~\ref{thm_static} with the following.
	
	\begin{lemma} \label{lemma_static}
		$\spec{UPD}^{\Psi}(\vec{Y})$ is statically equivalent to $\impl{UPD}^{\Psi, \Omega}(\vec{Y})$. 
		\begin{proof}
			Considering the definition of $\mathfrak{R}$ in Fig.~\ref{fig_relation}, let $ \nw \vec{x}.\rounds{\sigma \cpar P} \triangleeq \spec{UPD}^{\Psi}(\vec{Y})$ and $\nw \vec{y}.\rounds{\theta \cpar Q} \triangleeq \impl{UPD}^{\Psi, \Omega}(\vec{Y})$. We aim to show that  $\nw \vec{x}.\rounds{\sigma \cpar P}$ is statically equivalent to $\nw \vec{y}.\rounds{\theta \cpar Q}$. Since $\vec{x}$ is always a superset of $\vec{y}$, we prove that for all messages $M$ and $N$, s.t. $\vec{x} \text{ \fresh{} } M, N$, we have $M\sigma =_E N\sigma$ if and only if $M\theta =_E N\theta$.
			
			Recall the definition of $m^d(a, y)$:
			\begin{equation}\notag
				\mbox{$\squeezespaces{0.00001}m^d(a, y) \coloneqq \enc{\pair{\smult{a}{\pk{c_d}}}{\smult{a}{\sig{s}{\pk{c_d}}}}}{\hash{\smult{\mult{a}{c_d}}{y}}}$} 
			\end{equation}
			
			Since it is sufficient to consider normalisations when proving static equivalence, we present the normalisations for a fixed partitions $\{\alpha, \beta, \gamma, \delta\}$, $\{\zeta^1, \hdots, \zeta^D\}$ of the set of all sessions $\{1, \hdots, L\}$ of $\sigma$ and $\theta$ with respect to $E$ below. 
			
			\begin{equation} \notag
				\begin{array}{ll}
					pk_s \sigma = \pks{s}
					\\
					u_l \sigma = ch_l \mbox{ if $l \in \{1, \hdots, L\}$}
					\\
					v_l \sigma = \smult{\mult{a_l}{c_l}}{\gen} \mbox{ if $l \in \beta \cup \gamma \cup \delta$}
					\\[2pt]
					\mbox{if $l \in \delta$ and $Y_l = \smult{T_l}{\gen}$}
					\\
					\proj{\dec{\hash{\smult{T_l}{v_l}}}{w_l}} \sigma = \smult{\mult{a_l}{c_l}}{\gen}
					\\
					\projj{\dec{\hash{\smult{T_l}{v_l}}}{w_l}} \sigma = \smult{\mult{a_l}{c_l}}{\sig{s}{\gen}}
					\\
					\mbox{$\squeezespaces{0.7}\checksig{pk_s}{\projj{\dec{\hash{\smult{T_l}{v_l}}}{w_l}}} \sigma = \smult{\mult{a_l}{c_l}}{\gen}$}
					\\[2pt]
					\mbox{if $l \in \delta$ and $Y_l \neq \smult{T_l}{\gen}$}
					\\
					w_l \sigma = m^l(a_l, Y_l \sigma)
					\\[8pt]
					pk_s \theta = \pks{s}
					\\
					u_l \theta = ch_l \mbox{ if $l \in \{1, \hdots, L\}$}
					\\
					v_l \theta = \smult{\mult{a_l}{c_d}}{\gen} \mbox{ if $l \in \zeta^d \cap \left(\beta \cup \gamma \cap \delta \right)$}
					\\[2pt]
					\mbox{if $l \in \zeta^d \cap \delta$ and $Y_l = \smult{T_l}{\gen}$}
					\\
					\proj{\dec{\hash{\smult{T_l}{v_l}}}{w_l}} \theta = \smult{\mult{a_l}{c_d}}{\gen}
					\\
					\projj{\dec{\hash{\smult{T_l}{v_l}}}{w_l}} \theta = \smult{\mult{a_l}{c_d}}{\sig{s}{\gen}}
					\\
					\mbox{$\squeezespaces{0.7}\checksig{pk_s}{\projj{\dec{\hash{\smult{T_l}{v_l}}}{w_l}}} \theta = \smult{\mult{a_l}{c_d}}{\gen}$}
					\\[2pt]
					\mbox{if $l \in \zeta^d \cap \delta$ and $Y_l \neq \smult{T_l}{\gen}$}
					\\
					w_l \theta = m^d(a_l, Y_l \theta)
				\end{array}
			\end{equation} 
			
			We prove static equivalence by induction on the structure of the weak
			normal form of $N\sigma$ exploring all cases allowed by the grammar
			in Fig.~\ref{fig_syntax}. We present proofs starting from the equation under the frame $\sigma$. The argument for the converse case is the same. From now on $M$, $M_k$, $N$, $N_k$ are always fresh for $\vec{x}$.
			
			\textit{Case} 1. $N\sigma =_E \gen$. 
			
			\textit{Case} 1.1. $N = \gen$. If $M$ is a recipe for $\gen$, 
			then $M = \gen$, since there is no non-trivial recipe for $\gen$ under the normalisation of $\sigma$. Then we have $\gen\sigma =_E \gen\sigma$ if and only if $\gen\theta =_E \gen\theta$ as required. 
			
			\textit{Case} 1.2. $N \neq \gen$. There is nothing to prove in this case, since there is no non-trivial recipe for $\gen$ under the normalisation of $\sigma$. 
			
			\textit{Case} 2. $N\sigma =_E z$, $z$ is a variable. 
			
			\textit{Case} 2.1. $N = z$. If $M$ is a recipe for $z$, 
			then $M = z$, since there is no non-trivial recipe for $z$ under the normalisation of $\sigma$. Then we have $z\sigma =_E z\sigma$ if and only if $z\theta =_E z\theta$ as required. 
			
			\textit{Case} 2.2. $N\sigma =_E ch_l$. Since $N$ is fresh for $\vec{x}$, $N = u_l$. There is unique recipe $M = u_l$ for $ch_l$ and we have $u_l\sigma =_E u_l\sigma$ if and only if $u_l\theta =_E u_l\theta$ as required.  
			
			\textit{Case} 3. $N\sigma = \mult{K_1}{K_2}$.
			
			Any message term in the normalisation of $\sigma$ is $m$-atomic, hence no message is an immediate $m$-factor of another message in the normalisation of $\sigma$. Therefore there is only one case to consider. 
			
			\textit{Case} 3.1. $N = N_1^{j_1} \cdot \, \hdots \, \cdot
			N_k^{j_k}$, that is $N\sigma$ is generated by $m$-factors which have
			a recipe under the normalisation of $\sigma$: $N\sigma =
			N_1^{j_1}\sigma \cdot \, \hdots \, \cdot N_k^{j_k}\sigma$. By the
			induction hypothesis suppose that for all recipes $M_i$ for an
			$m$-factor $N_i\sigma$ of $N\sigma$, we have $M_i\sigma =_E
			N_i\sigma$ if and only if $M_i\theta =_E N_i\theta$, $i \in \{1,
			\hdots, k\}$. By applying multiplication, we have $M_1^{j_1}\theta
			\cdot \, \hdots \, \cdot M_k^{j_k}\theta = (M_1^{j_1} \cdot \, \hdots
			\,\cdot M_k^{j_k})\theta =_E (N_1^{j_1} \cdot \, \hdots \, \cdot
			N_k^{j_k})\theta = N_1^{j_1}\theta \cdot \, \hdots \, \cdot
			N_k^{j_k}\theta$ as required, and $N_i\theta$ is an $m$-factor of
			$N\theta$.
			
			\textit{Case} 4. $N\sigma = \smult{K_1}{K_2}$.
			
			Let us define 
			\begin{equation}\notag
				\begin{array}{ll}
					V_1 \coloneqq v_l, \,V_2 \coloneqq \proj{\dec{\hash{\smult{T_l}{v_l}}}{w_l}}, \\[2pt]
					V_3 \coloneqq \projj{\dec{\hash{\smult{T_l}{v_l}}}{w_l}} \\[2pt]
					V_4 \coloneqq  \checksig{pk_s}{\projj{\dec{\hash{\smult{T_l}{v_l}}}{w_l}}}
				\end{array}
			\end{equation}
			\textit{Case} 4.1. $N\sigma = \smult{\mult{a_l}{c_l}}{\gen}$ and $Y_l =  \smult{T_l}{\gen}$. Since $N$ is fresh for $\vec{x}$, $N \in \{V_1, V_2, V_4\}$. Let $M$ be a recipe for $\smult{\mult{a_l}{c_l}}{\gen}$, then $M \in \{V_1, V_2, V_4\}$ and we have $M\sigma =_E N\sigma$ if and only if $M\theta =_E N\theta$ for any $N$ and $M$ as required. In case $Y_l \neq \smult{T_l}{\gen}$, $N=V_1$, there is unique recipe $M_1=V_1$ and the argument is the same. 
			
			\textit{Case} 4.2. $N\sigma = \smult{\mult{a_l}{c_l}}{\sig{s}{\gen}}$ and $Y_l = \smult{T_l}{\gen}$. Since $N$ is fresh for $\vec{x}$, $N=V_3$ and there is unique recipe $M = V_3$ for $\smult{\mult{a_l}{c_l}}{\gen}$, and we have $V_3\sigma =_E V_3 \sigma$ if and only if $V_3 \theta =_E V_3 \theta$ as required. If $Y_l \neq \smult{T_l}{\gen}$, there is no recipe for $\smult{\mult{a_l}{c_l}}{\sig{s}{\gen}}$ and there is nothing to prove. 
			
			\textit{Case} 4.3. $N = \smult{N_1}{N_2}$, $N_2 \in \{V_1, V_2, V_3, V_4\}$ and $Y_l =  \smult{T_l}{\gen}$. By the induction hypothesis  suppose that for all recipes $M_1$ for $N_1\sigma$ , we have $M_1\sigma =_E N_1\sigma$ if and only if $M_1\theta =_E N_1\theta$, then multiply $N_2$ by a scalar $M_1$ and obtain $\smult{M_1\theta}{N_2\theta} = \smult{M_1}{N_2}\theta =_E \smult{N_1}{N_2}\theta = \smult{N_1\theta}{N_2\theta}$ for any $N_2$ as required. In case $Y_l \neq \smult{T_l}{\gen}$, $N_2 = V_1$ and the argument is the same. 
			
			\textit{Case} 4.4. $N =\texttt{sig}(\hdots \, \sig{N_2}{N_1} \, \hdots \, ,N_k)$, $N_1 \in \{V_1, V_2, V_3, V_4\}$ and $Y_l =  \smult{T_l}{\gen}$. By the induction hypothesis suppose that for all recipes $M_i$ for $N_i\sigma$ we have $M_i\sigma =_E N_i\sigma$ if and only if $M_i\theta =_E N_i\theta$, $i \in \{2, \hdots, k\}$. By applying the signature operation to $N_1$, we have 
			\begin{equation}\notag
				\begin{array}{ll}
					\texttt{sig}(\hdots \, \sig{M_2}{N_1} \, \hdots \, ,M_k)\theta = \\[2pt]
					\texttt{sig}(\hdots \, \sig{M_2\theta}{N_1\theta} \, \hdots \, ,M_k\theta) =_E \\[2pt]
					\texttt{sig}(\hdots \, \sig{N_2\theta}{N_1\theta} \, \hdots \, ,N_k\theta) = \\[2pt]
					\texttt{sig}(\hdots \, \sig{N_2}{N_1} \, \hdots \, ,N_k)\theta
				\end{array}
			\end{equation}
			as required. In case $Y_l \neq \smult{T_l}{\gen}$, $N_1 = V_1$ and the argument is the same. 
			
			\textit{Case} 4.5. $N = \smult{N_1}{N_2}$. Similar to case 3.1 with $j_i=1, i\in\{1, 2\}$. 
			
			\textit{Case} 5. $N\sigma = \pair{K_1}{K_2}$.
			
			Since no pair is contained in the normalisation of $\sigma$, there is only one case to consider. 
			
			\text{Case} 5.1. $N = \pair{N_1}{N_2}$. Similar to case 4.5.
			
			\textit{Case} 6. $N\sigma = \hash{K_1}$. Similar to case 5 with $j_i=1, i=1$.
			
			\textit{Case} 7. $N\sigma = \pks{K_1}$. 
			
			\textit{Case} 7.1. $N\sigma = \pks{s}$. Then $N=pk_s$, since $N$ is fresh for $\vec{x}$. There is a unique recipe $M = pk_s$ for $\pks{s}$ and we have $pk_s\sigma =_E pk_s\sigma$ if and only if $pk_s\theta =_E pk_s\theta$ as required.
			
			\textit{Case} 7.2. $N = \pks{N_1}$. Similar to case 6.
			
			\textit{Case} 8. $N\sigma = \sig{K_2}{K_1}$. Similar to case 5.
			
			\textit{Case} 9. $N\sigma = \enc{K_1}{K_2}$. 
			
			If $Y_l = \smult{T_l}{\gen}$ no encrypted 
			message term is contained in the normalised frame and there is only one case to consider. 
			
			\textit{Case} 9.1. $N = \enc{N_1}{N_2}$. Similar to case 5.
			
			If $Y_l \neq \smult{T_l}{\gen}$, there is also the following.
			
			\textit{Case} 9.2. $N\sigma =_E m^l(a_l, Y_l \sigma)$. Since $N$ is fresh for $\vec{x}$, $N = w_l$. There is unique recipe $M = w_l$ for $m^l(a_l, Y_l \sigma)$ and we have $w_l\sigma = w_l\sigma$ if and only if $w_l\theta = w_l\theta$ as required.
			
		\end{proof}
	\end{lemma}
	
	\section{Unlinkable Authentication for BDH}\label{sec_auth}
	
	The twofold aim of the BDH protocol is to guarantee unlinkability of the card,
	while allowing the terminal to authenticate the card.
	In this paper we emphasise unlinkability, since this is the more novel of the two requirements.
	Indeed, ProVerif, and other tools, can be used to automatically confirm our target authentication property -- injective agreement~\cite{lowe1997auth} -- holds for both BDH protocols in this paper.
	
	The process scheme below specifies the behaviour of honest terminals and honest cards.
	The attacker is the implicit environment that interacts with these honest participants.
	
	\[
	\System \triangleeq
	\begin{array}[t]{l}
		\nw s.\Big(
		\arraycolsep=0pt
		\begin{array}[t]{l}
			\bang\nw c.\bang
			\nu ch_c. \cout{\cch}{ch_c}.\card(s, c, ch_c)   ~\cpar                
			\\
			\cout{out}{\pks{s}}.       \bang \nu ch_t. \cout{\tch}{ch_t}.\terminal(\pks{s}, ch_t) ~\Big)
		\end{array}
	\end{array}
	\]
	
	In the above, the processes $\card$ and $\terminal$ can be instantiated with $\bdh{\card}$ and $\bdh{\terminal}$
	or with $\fix{\card}$ and $\fix{\terminal}$ to obtain $\bdh{\System}$
	and $\fix{\System}$, respectively.
	Notice a fresh channel for each run is advertised on channels $\cch$ or $\tch$.
	These allow the messages associated with a run to be uniquely identified in the formulation of injective agreement below.
	
	The following injective agreement property is standard~\cite{lowe1997auth,cremersmauw}.
	\textit{Agreement} here means that when a terminal thinks it has authenticated a card, an honest card really executed the protocol while exchanging the same messages as the terminal.
	\textit{Injectivity} strengthens agreement by ensuring that every successfully authenticating run of a terminal corresponds to a separate run of a card.
	\begin{definition}(injective agreement) Process $\System$ satisfies injective agreement
		iff
		for every trace
		$\pi_0$, $\pi_1$, \ldots, ${\pi_n}$
		such that
		$
		\System \vDash \diam{\pi_0} \ldots \diam{\pi_n} \texttt{true}
		$\footnote{We reuse our modal logic to describe traces satisfied by a process,
			extended such that $A \vDash \texttt{true}$ holds, and also $A \vDash \phi \wedge \psi$ iff $A \vDash \phi$ and $A \vDash \psi$.
		}
		there exists an injective function $f\colon  \mathbb{N} \rightarrow \mathbb{N}$ 
		such that,
		for every $a$ such that $0 < a \leq n$, $\pi_a = \co{T_a}(w)$ and 
		$
		\System \vDash \diam{\pi_0} \ldots \diam{\pi_n}( w = \ack )
		$,
		we have the following:
		\begin{itemize}
			\item
			for some $0 \leq i < j < k < a$,
			we have the following
			$\pi_i = {T_i}\,{M_i}$,
			$\pi_j = \co{T_j}(u_j)$,
			and $\pi_k = {T_k}\,{M_k}$;
			
			\item
			for $0 \leq f(a) < i' < j' < k' < a$,
			s.t.~
			$\pi_{f(a)} = \co{C_{f(a)}}{(ch_c)}$,
			we have
			$\pi_{i'} = \co{C_{i}}(u_i)$,
			$\pi_{j'} = {C_{j}}\,{M_j}$,
			and $\pi_{k'} = \co{C_{k}}(u_k)$,
			
			\item
			and
			$
			\System \vDash \diam{\pi_0} \ldots \diam{\pi_n}\left( 
			C_{f(a)} = \cch
			\wedge
			\phi_i
			\wedge
			\phi_j
			\wedge
			\phi_k
			\right)
			$,
			where $\phi_\ell 
			\triangleq
			u_\ell = M_\ell
			\wedge
			ch_c = C_\ell
			\wedge
			T_a = T_\ell$.
		\end{itemize}
	\end{definition}
	
	We can now verify that our target functional property holds.
	The proof is conducted in ProVerif [see Appendix~\ref{app_proverif} for the code], with 
	respect to an extension of the standard Diffie-Hellman theory for ProVerif,
	which approximates the equations for multiplication
	with the equation $\smult{a}{\smult{b}{\gen}} = \smult{b}{\smult{a}{\gen}}$.
	We had to extend that standard theory further with an equation
	$\smult{a}{\smult{b}{\smult{c}{\gen}}} = \smult{b}{\smult{a}{\smult{c}{\gen}}}$, so that blinding factors are treated correctly.
	\begin{theorem}\label{thm_auth}
		\begin{upshape}$\bdh{\System}$\end{upshape} and \begin{upshape}$\fix{\System}$\end{upshape} satisfy injective agreement.
	\end{theorem}\
	Despite ProVerif requiring an approximation of the Diffie-Hellman theory, we find this proof to be sufficient, since authentication already held for the BDH protocol of EMVCo, and we simply aim to show that our proposed fix does not inadvertently break authentication. Notice in particular that our attacker is incapable of relating two public keys, i.e. if she knows $pk_1 \coloneqq \pk{c_1}$ and $pk_2 \coloneqq \pk{c_2}$, it is infeasible for her to find a scalar $h$ s.t. $\mult{h}{pk_1}=pk_2$.
	This contrasts, to our thorough proof of unlinkability (Theorem~\ref{thm_static}), which takes equations in Fig.~\ref{fig_eqt} and Fig.~\ref{fig_blinding} fully into account.
	
	\section{Unlinkability challenges in EMV 1st Gen}\label{sec_discussion}
	
	In this section we explain that even in the presence of the UBDH key agreement it is challenging to make an entire EMV 1st Gen transaction unlinkable without substantial changes to the current protocol~\cite{emv} and the back-end, i.e., bank-terminal communications. 
	Notice that EMVCo never has made precise how BDH and the rest of the
	current EMV protocol coexist. In this section we assume the scenario
	in which the key agreement protocol is run before any data about the
	transaction is transmitted and we informally describe a generic EMV transaction highlighting steps where identifying information about the card is revealed to the terminal. 
	In this discussion we account for the following sources of identities the card has: unique identifiers such as the card number; and coarser identifiers such as the data formats that the card supports. We warn the reader that this section is not a comprehensive EMV protocol description. The current EMV standard admits optional steps that are up to a particular payment system to implement. For improved clarity, these optional steps are mostly omitted. 
	
	\subsection{EMV transaction flow}
	
\begin{figure}[h]
		\includegraphics[width=\linewidth]{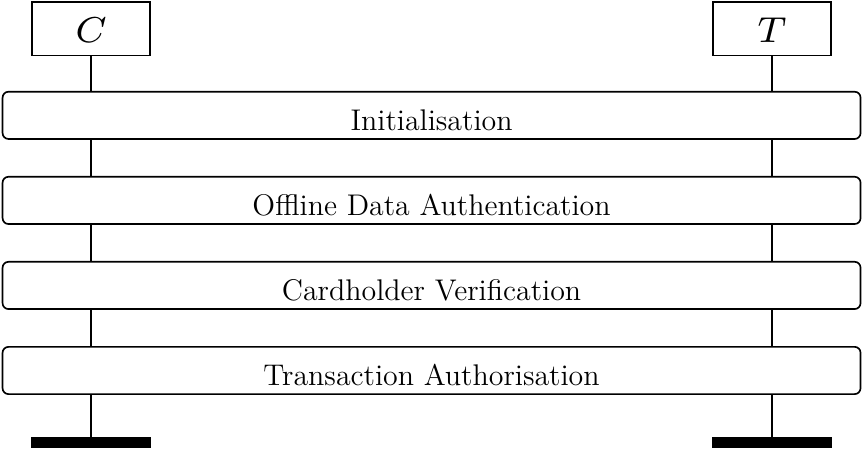}
		\centering
		\caption{The EMV 1st Gen protocol stages.} 
		\label{fig_emv}
	\end{figure}

	An EMV transaction is a sequence of the terminal's commands with an optional data payload and the card's responses. This message exchange is broken down into several stages presented in Fig.~\ref{fig_emv}. A successful transaction ends with the generation of a cryptogram that the terminal eventually sends to the bank in exchange for money. In what follows, we briefly discuss each stage and summarise which of the unlinkability issues are relevant for the eavesdropping-resistant and the active attacker-resistant models.
	
	\subsubsection{Initialisation} 
	
	Firstly, the terminal asks the card which applications it supports.
	The card responds with a list of payment application identifiers, e.g. Visa Debit, Maestro, etc. Then the terminal selects a particular application from the presented list.
	If the intention of EMVCo is that BHD is run immediately after this step to protect further communication, a passive eavesdropper can distinguish two cards selecting different applications thereby violating a coarse form of external unlinkability. This could be averted by all cards having the same identifier, and, in addition, if the key agreement were further enhanced by group signatures. A group signature allows all cards to be signed with different keys, and verified with a single key, without different card issuers revealing their keys to each other. This is an example of something that could be reviewed by the developers of the EMV 2nd Gen standard.
	
	Having selected the application, the card sends the 
	PDOL list specifying which transaction details (e.g. the amount, the date, the currency, etc.) the terminal should send to the card.
	Next, in one message, the terminal sends the requested PDOL data and requests the AIP list, specifying the functions supported by the card including authentication methods and the AFL list, specifying memory addresses where the card stores its data such as the card number, the certificate, etc. Finally, the terminal uses the addresses from the AFL list to read the actual data from the card. The following data is mandatory for the card to have as specified in the EMV 1st Gen standard.
	\begin{itemize}
		\item Application Expiry Date
		\item The card's number.
		\item Card Risk Management Data Object Lists (CDOLs)
	\end{itemize}
	CDOLs specify the information the card wants to receive from the terminal to generate cryptogram(s) at the end of the transaction, e.g. the country code, the terminal nonce, etc. Typically, the certificate and the payment system public key index is also available for the terminal to read. 
	
	We can see that the application selection step already violates
	unlinkability. As mentioned above, the list of supported payment
	systems serves a coarse form of identity, but, fortunately, the PDOL
	and the transaction data it specifies would be protected from a
	passive eavesdropper by running BDH. However, these messages would not
	be protected from an active attacker, and hence would be available to
	distinguish cards. The same applies to the AIP, AFL, CDOL lists, and the public key index. All this information varies from card to card and contributes to the card's fingerprint that can be used by an attacker to link sessions with the same card with high probability. 
	Notice that it is relatively easy to remove certain messages so they do not contribute to the fingerprint of the card. For instance, all cards may request a standard set of transaction details from the terminal making the PDOL obsolete. The others, such as the list of supported payment applications, would require a substantial reworking of the current protocol. Such reworking would be even more challenging since it should account for the strong forms of the card's identity such as the expiry date, the card's number, and the certificate. 
	The most difficult point to address in order to achieve strong unlinkability, is how to hide the card number from the terminal, since, in EMV 1st Gen, it is revealed in order to route the cryptogram through the network to the bank at the last stage of the transaction.
	
	\subsubsection{Offline data authentication}
	
	Offline Data Authentication is an optional step in the EMV 1st Gen
	protocol at which point the terminal authenticates the data previously
	received from the card during the Initialisation step. The ODA can be
	completed in several ways.\footnote{\scriptsize For backward compatibility, EMV retains also a Static Data Authentication (SDA) mode, but, this should be avoided in new cards, since it facilitates card cloning by replaying the signature~\cite{van2016emv}.}
	
	\begin{itemize}
		\item Dynamic Data Authentication (DDA). At the Initialisation
		phase, the card could provide the DDOL with the list of data elements that the terminal must send to the card if DDA is selected. This DDOL is a coarse identity that contributes to the card's fingerprint. In case the DDOL is not provided, the terminal should always send a nonce to the card. 
		The card replies with a signature on the DDOL data provided by the terminal and its dynamic data (e.g. a nonce generated by the card). This signature itself does not reveal identifying information about the card, but to verify it the terminal uses the card's public key obtained from the certificate, both of which are card identifiers.
		\item Combined Data Authentication (CDA). This case does not require additional messages and is a part of the Transaction Authorisation phase. Otherwise, it is similar to DDA, but the card's dynamic data also includes transaction details, e.g. the cryptogram.
	\end{itemize}
	
	A selection of a particular Offline Data Authentication method itself does not contribute to the card’s fingerprint since the supported methods have already been disclosed at the Initialisation step by providing the terminal with the AIP, however, we can see that all three Offline Data Authentication methods involve a unique card identifier that, if exposed to a terminal, break unlinkability with respect to an active attacker. For a passive eavesdropper, the data communicated in this phase is unavailable since she is already locked out of the session. In the presence of either BDH or UBDH, the ODA phase is built in to the key agreement, hence could be redundant in a EMV 2nd Gen protocol.

	\subsubsection{Cardholder Verification}
	
	Cardholder Verification is also optional in the current EMV protocol. Cards supporting Cardholder Verification provide a list of supported methods in the Initialisation step. This list is a coarse form of the card's identity, contributes to its fingerprint, and is available only to active attackers. Verification methods include a handwritten signature, PIN, and a verification via the consumer's device (e.g. through biometric data entered via a mobile phone) which is out of the scope of the EMV standard. For any threat model we consider in this paper we have to assume that a signature or PIN may only be disclosed to an honest party. For example, a PIN is never entered into a dishonest terminal, because otherwise, if a malicious terminal learns the PIN, the basic security requirement of EMV, the safety of money in the cardholder's account, is compromised; not only the privacy.

	\subsubsection{Transaction Authorisation}
	
	Transaction Authorisation is the ultimate and mandatory phase 
	at which the terminal asks the card to generate the Application Cryptogram (AC). A cryptogram is typically a Hash-based Message Authentication Code (HMAC) generated by the card over the data coming both from the card and the terminal (specified by CDOLs the terminal has received in the Initialisation phase) using the key derived from the shared secret $mk$ between the card and the bank and the ATC (Application Transaction Counter). The official EMVCo recommendations on the minimum set of data elements to be included in the cryptogram are listed in Book 2 of the EMV 1st Gen standard~\cite{emv} and include, for instance, the type of the cryptogram (decline, approve, request online), the card number, ATC, etc. Notice that together with the cryptogram, the data that it was generated over must be provided to verify this cryptogram, which causes linkability issues in the presence of active attackers, i.e. the value of the ATC and the card number are forms of the card identity. 
	
	\subsection{The future of Unlinkable EMV transactions}
	
	The above discussion demonstrates that even the promotion of the anti-eavesdropping requirement to full unlinkability in the presence of \emph{passive} attackers, relevant in the range from 1m to 20m, 
	demands updates to the EMV 1st Gen, e.g. at the application selection step. This potential minor update would be impossible to roll out incrementally since it would require a great cooperation effort between EMVCo and adopters. For a major update, that strengthens the requirements further to support unlinkability with respect to \emph{active} attackers, relevant within 1m from the card, our analysis clearly suggests that it is unfeasible to hide all identifying information from the terminal without touching critical elements of the standard. Making the messages that contribute to the fingerprint of the card constant and, hence, obsolete would reduce the flexibility of EMV. The direct card's identifiers play a crucial role in steps that ensure the primary EMV goal, a safe money exchange: e.g. the card number is important for network routing, the card's public key is involved in data authentication, etc. 
	
	We conclude that the anti-eavesdropping requirement, more specifically, anti-eavesdropping on the phase where the transaction data is communicated, is the only thing BDH fulfills. UBDH, on the other hand, targets a much more ambitious goal, and the respective updates would inevitably require larger compromises and infrastructure updates. Whether such an update to the full transaction protocol exists is future work that should be addressed in coordination with EMVCo.

	\section{Conclusion}\label{sec_concl}
	
	In this paper, we have investigated the Blinded Diffie-Hellman key
	agreement protocol in Fig.~\ref{fig_orig} proposed by EMVCo to
	introduce encryption into a proposal for 2nd Gen EMV
	payments. Although BDH indeed introduces a way to establish a
	symmetric key between the card and the terminal, and meets the
	initial EMVCo requirements, we have shown that the privacy of the
	cardholder will not be protected against an active attacker. In particular, in
	Theorem~\ref{thm_link} we have shown that the presence of an active
	adversary leads to a straightforward failure of BDH to be unlinkable. 
	In our proposal for improving the protocol in Fig.~\ref{fig_impr}, we use a generic signature scheme that
	respects blinding. To support this, in Section~\ref{sec_fixed}, we point out that at least one existing signature scheme meets our requirements, namely Verheul signatures. To verify our proposal, we introduced a strong definition of
	unlinkability in Def.~\ref{def_unlink} and applied this definition to
	the applied $\pi$-calculus model of UBDH in Theorem~\ref{thm_static},
	thereby proving that UBDH indeed makes the key agreement unlinkable.
	
	The first, core, take away message of our paper is related to the threat model in the EMVCo proposal of secure channel
	establishment~\cite{rfc}. The anti-eavesdropping requirement in the BDH and 2nd Gen specifications~\cite{rfc, overview} form a reasonable privacy enhancement
	guided by the current state of the EMV standard and the infrastructure
	already deployed. We, however, have taken the liberty to look beyond
	passive eavesdropping and considered the implications of realistic active attackers,
	as captured in Def.~\ref{def_unlink}. We support this investigation by observing that, the anti-tracking requirement explicitly mentioned in the BDH proposal remains open to interpretation, specifically~\cite{rfc}, ``The protocol is designed to protect against eavesdropping and card tracking.'' We conclude that unlinkable EMV key agreement, under such assumptions, is feasible as we prove formally in Theorem~\ref{thm_link}. Yet, unlinkability in the presence of an active attacker is
	difficult to extend to the full EMV 1st Gen transaction as we discuss
	informally in Section~\ref{sec_discussion}.

	The second, more general, take away message concerns our method for verifying properties defined in terms of a process equivalence problem. Our evolved unlinkability definition, Def.~\ref{def_unlink}, based on the notion of quasi-open bisimilarity, which is a congruence,  enables compositional reasoning about protocols, such as UBDH, without a shared key. This state-of-the-art approach to bisimilarity checking facilitates proving that the property holds for unboundedly many sessions, where the main challenge is to define the relation in Fig.~\ref{fig_relation}, after which we apply the method to show that the relation is a quasi-open bisimulation. Furthermore, the equational theory we employ is not yet covered by equivalence checking tools, so this example proof may help inform the extension of tools to this class of problems.
	
	EMVCo is still in the process of revising the protocol for the EMV 2nd Gen standard~\cite{emv2019statement}. As awareness of these privacy issues is growing and methods, such as ours, for verification of privacy properties emerging, we expect stakeholders to take seriously the possibility of making payments unlinkable. 
	
	\paragraph*{Acknowledgements}
	We thank the anonymous reviewers for their constructive suggestions.
	
	\bibliography{ref}{}
	\bibliographystyle{myIEEEtran}
	
	\appendix 
	
	\subsection{Background on normalisation used in the proof of Lemma~\ref{lemma_static}}\label{app_norm}
	The \emph{weak normal form} $\nf{M}$ of a term $M$ with respect to $E$ captures the least complex (up to multiplication) expression of $M$. We do not require the weak normal form to be unique.
	\begin{definition}(weak normal form) The weak normal $\nf{M}$ of a message term $M$ is defined inductively on the structure of $M$:
		\begin{itemize}
			\item $M=\gen$ or $M$ is a variable, then $\nf{M} = M$.
			\item $M= \mult{M_1}{M_2}$, then $\nf{M} = \mult{\nf{M_1}}{\nf{M_2}}$.
			\item $M = \smult{M_1}{M_2}$, then $\nf{M} = \smult{\nf{M_1}}{\nf{M_2}}$ if $\nf{M_2}$ is $\phi$-atomic. Otherwise $\nf{M} = \smult{\mult{\nf{M_1}}{\nf{M_2'}}}{\nf{M_2''}}$, where $M_2 =_E \smult{M_2'}{M_2''}$ and $\nf{M_2''}$ is $\phi$-atomic.
			\item $M=\pair{M_1}{M_2}$, then $\nf{M} = \pair{\nf{M_1}}{\nf{M_2}}$.
			\item $M=\hash{M_1}$ or $M=\pks{M_1}$, then $\nf{M} = \hash{\nf{M_1}}$ or $\nf{M} = \pks{\nf{M_1}}$ respectively. 
			\item $M = \sig{M_2}{M_1}$, then $\nf{M} = \sig{\nf{M_2}}{\nf{M_1}}$ if $\nf{M_1}$ is $\phi$-atomic. Otherwise $\nf{M} = \smult{\nf{M_1'}}{\sig{\nf{M_2}}{\nf{M_1''}}}$, where $M_1 = \smult{M_1'}{M_1''}$ and $\nf{M_1''}$ is $\phi$-atomic.
			\item $M = \proj{\pair{M_1}{M_2}}$ or $M = \projj{\pair{M_1}{M_2}}$ then $\nf{M} = \nf{M_1}$ or $\nf{M} = \nf{M_2}$ respectively.
			\item $M = \dec{M_2}{\enc{M_1}{M_2}}$, then $\nf{M} = \nf{M_1}$.
			\item $\squeezespaces{0.83}M = \checksig{\pks{M_2}}{\sig{M_2}{M_1}}$, then $\squeezespaces{0.82}\nf{M} = \nf{M_1}$.
			\item Otherwise $\nf{M}=M$.
		\end{itemize}
	\end{definition}
	For a process $\nw \vec{x}.\rounds{\sigma \cpar P}$, the \emph{normalisation} of $\sigma$ is a frame with recipes allowed in the domain constructed as follows.
	\begin{enumerate}
		\item $u\sigma=M$ for any $u \in \dom{\sigma}$ is replaced by $u\sigma = \nf{M}$.
		\item If $u\sigma = \mult{K_1}{K_2}$ and there is a recipe $M_1$ for an immediate $m$-factor $K_1$, then $M_1\sigma$ is added to the normalisation. If there is a recipe $M_2$ for an immediate $m$-factor $K_2$, then $M_2\sigma$ is also added to the normalisation. 
		\item If $u\sigma = \pair{K_1}{K_2}$, then $u\sigma$ is replaced by $\proj{u}\sigma = K_1$ and $\projj{u}\sigma = K_2$.
		\item If $u\sigma = \enc{K_1}{K_2}$ and there is a recipe $M_2$ for $K_2$, then $u\sigma$ is replaced by $\dec{M_2}{u}\sigma = K_1$.
		\item If $u\sigma = \sig{N_2}{N_1}$ and there is a recipe $M_2$ for $N_2$, then $u\sigma$ is replaced by $\checksig{\pks{M_2}}{u}\sigma = N_1$. 
		\item If $u\sigma = \sig{N_2}{N_1}$ and there is a recipe $M_2$ for $\pks{N_2}$, then $\checksig{M_2}{u}\sigma = N_1$ is added to the normalisation.
	\end{enumerate}
	
	To give an example, consider the extended process $$\squeezespaces{0.1}\nw \vec{x}.(\sigma \cpar P) \triangleeq \nw s, a, b. \left(\sub{pk_s, u_1, u_2, u_3}{\pks{s}, \enc{\hash{a}}{b}, b, \sig{s}{\pair{a}{x}}}\cpar P \right)$$
	Then the normalisation of $\sigma$ is given below.
	\begin{equation}\notag
		\begin{array}{ll}
			pk_s \sigma = \pks{s}
			\\
			\dec{u_2}{u_1} \sigma = \hash{a}
			\\
			u_2 \sigma = b
			\\
			\proj{\checksig{pk_s}{u_3}}\sigma = a
			\\
			\projj{\checksig{pk_s}{u_3}}\sigma = x
			\\
			u_3 \sigma = \sig{s}{\pair{a}{x}}
		\end{array}
	\end{equation} 
	
	The advantage of working with normalisations is the reduction in message complexity in the range of $\sigma$ without affecting static equivalence: $M$ is a recipe under $\sigma$ if and only if $M$ is a recipe under the normalisation of $\sigma$. 
	
	\subsection{ProvVerif code supporting the proof of Theorem~\ref{thm_auth}}\label{app_proverif}

\vspace{10pt}

BDH satisfies injective agreement.

\vspace{10pt}

\normalsize
\begin{lstlisting}
free cout, card, term: channel.

type key.
type sskey.
type spkey.
type point.
type scalar.

fun smult(scalar, point): point.
fun h(point): key.
fun pk(sskey): spkey.
fun sign(point , sskey): bitstring.
fun enc(bitstring, key): bitstring.

const G: point [data].

reduc forall m: bitstring, k: key; 
 dec(enc(m, k), k) = m.
reduc forall m: point, k: sskey; 
 check(sign(m, k), pk(k)) = m.

equation forall a: scalar, b: scalar; 
 smult(a, smult(b, G)) = 
 smult(b, smult(a, G)).
equation forall a: scalar, b: scalar, 
 c: scalar; 
 smult(a, smult(b, smult(c, G))) = 
 smult(b, smult(a, smult(c, G))).

event snd1(point).
event rec1(point).
event snd2(point).
event rec2(point).
event cardTerm(key, bitstring).
event terminalTerm(key, bitstring).

(*check that terminalTerm is reachable*)
query k: key, z2: bitstring;
 event(terminalTerm(k, z2)).

query k: key, z2: bitstring, z1: point;
 inj-event(terminalTerm(k, z2)) ==> 
 (inj-event(rec1(z1)) ==> 
 inj-event(snd1(z1))).

query k: key, z2: bitstring, y: point;
 inj-event(terminalTerm(k, z2)) ==> 
 (inj-event(rec2(y)) ==> 
 inj-event(snd2(y))).

query k: key, z2: bitstring;
 inj-event(terminalTerm(k, z2)) ==> 
 inj-event(cardTerm(k, z2)).

let C(s: sskey, c: scalar, ch: channel) =
 new a: scalar;
 event snd1(smult(a, smult(c, G)));
 out(ch, smult(a, smult(c, G)));
 in(ch, y: point);
 event rec2(y);  
 let k = h(smult(a, smult(c, y))) in
 let cert = (smult(c, G), 
  sign(smult(c, G), s)) in
 event cardTerm(k, enc(((a, smult(c, G)), 
  cert), k));
 out(ch, enc(((a, smult(c, G)), cert), k)).

let T(s: sskey, ch: channel) =
 new t: scalar;
 in(ch, z1: point);
 event rec1(z1);
 event snd2(smult(t, G));
 out(ch, smult(t, G));
 in(ch, z2: bitstring);
 let k = h(smult(t, z1)) in
 let ((n1: scalar , n2: point), 
  (n3: point, n4: bitstring)) = 
   dec(z2, k) in
 if (n2 = check(n4, pk(s))) && 
  (smult(n1, n2) = z1) then
 event terminalTerm(k, z2).

(*populate system with cards*)
let PopCard(s: sskey)=
 new c: scalar;
 !(new chc: channel;
 out(card, chc);
 C(s, c, chc)).

(*populate system with terminals*)
let PopTerminal(s: sskey)=
 new cht: channel;
 out(term, cht);
 T(s, cht).

process
 new s: sskey;
 out(cout, pk(s));
 !PopCard(s) | !PopTerminal(s)
\end{lstlisting}
\normalsize

\vspace{10pt}

UBDH satisfies injective agreement.

\vspace{10pt}

\normalsize
\begin{lstlisting}
free cout, card, term: channel.

type key.
type sskey.
type spkey.
type point.
type scalar.

fun smult(scalar, point): point.
fun h(point): key.
fun pk(sskey): spkey.
fun sign(point , sskey): point.
fun enc(bitstring, key): bitstring.

const G: point [data].

reduc forall m: bitstring, k: key; 
 dec(enc(m, k), k) = m.
reduc forall a: scalar, m: point, 
 k: sskey; 
 check(smult(a, sign(m, k)), pk(k)) = 
 smult(a, m).

equation forall a: scalar, b: scalar; 
 smult(a, smult(b, G)) = 
 smult(b, smult(a, G)).
equation forall a: scalar, b: scalar, 
 c: scalar; 
 smult(a, smult(b, smult(c, G))) = 
 smult(b, smult(a, smult(c, G))).


event snd1(point).
event rec1(point).
event snd2(point).
event rec2(point).
event cardTerm(key, bitstring).
event terminalTerm(key, bitstring).

(*check that terminalTerm is reachable*)
query k: key, z2: bitstring;
 event(terminalTerm(k, z2)).

query k: key, z2: bitstring, z1: point;
 inj-event(terminalTerm(k, z2)) ==> 
 (inj-event(rec1(z1)) ==> 
 inj-event(snd1(z1))).

query k: key, z2: bitstring, y: point;
 inj-event(terminalTerm(k, z2)) ==> 
 (inj-event(rec2(y)) ==> 
 inj-event(snd2(y))).

query k: key, z2: bitstring;
 inj-event(terminalTerm(k, z2)) ==> 
 inj-event(cardTerm(k, z2)).

let C(s: sskey, c: scalar, ch: channel) =
 new a: scalar;
 event snd1(smult(a, smult(c, G)));
 out(ch, smult(a, smult(c, G)));
 in(ch, y: point);
 event rec2(y);  
 let k = h(smult(a, smult(c, y))) in
 let m = (smult(a, smult(c, G)), 
  smult(a, sign(smult(c, G), s))) in
 event cardTerm(k, enc(m, k));
 out(ch, enc(m, k)).

let T(s: sskey, ch: channel) =
 new t: scalar;
 in(ch, z1: point);
 event rec1(z1);
 event snd2(smult(t, G));
 out(ch, smult(t, G));
 in(ch, z2: bitstring);
 let k = h(smult(t, z1)) in
 let (m1: point, m2: point) = 
  dec(z2, k) in
 if (m1 = check(m2, pk(s))) && 
  (m1 = z1) then
 event terminalTerm(k, z2).

(*populate system with cards*)
let PopCard(s: sskey)=
 new c: scalar;
 !(new chc: channel;
 out(card, chc);
 C(s, c, chc)).

(*populate system with terminals*)
let PopTerminal(s: sskey)=
 new cht: channel;
 out(term, cht);
 T(s, cht).

process
 new s: sskey;
 out(cout, pk(s));
 !PopCard(s) | !PopTerminal(s)
\end{lstlisting}
\normalsize

	\subsection{Proof trees for transitions in Theorem~\ref{thm_static}}\label{app_prooftrees}
	
   \vspace{10pt}
	
	By $\textsf{Rule}^n$ below we assume $n$ applications of the
	transition rule $\textsf{Rule}$ from Fig.~4. In case of $n$ consecutive applications of rules $\textsf{Par-l}$, $\textsf{Par-r}$ we write $\textsf{Par}^n$. Notice that $\alpha$-conversion is often used: in particular when the rule $\textsf{Extrusion}$ is applied. We define $\chi_l(\vec{Y}, M)$ as the list of message terms obtained by the replacement of $l$th entry in $\vec{Y}$ with $M$. In \textit{Case} 5, $\sigma'$, $\theta'$ are the frames accumulated at the point of input of $Y_l$. In the proof trees presented below we use the following abbreviations 
	
	\vspace{10pt}

	\begin{equation}\notag
		\begin{array}{ll}
			S \triangleeq \pspec(s, ch, c) \\[8pt] I \triangleeq \pimpl(s, ch, c)
		\end{array}
	\end{equation}
	
	
	\begin{figure*}[h]
		\[
		\scalebox{1}{
			\begin{prooftree}
				{
					\begin{prooftree}
						pk_s \text{ \fresh{} } out, s, \bang S \quad out \mathrel{=_{E}} out
						\justifies
						\mbox{$\squeezespaces{1}\cout{out}{\pks{s}}.\bang S \lts{\co{out}(pk_s)} \mathopen{\left(\sub{pk_s}{\pks{s}}\right) \cpar \bang S}$}
						\using  \mbox{\textsf{Out}}
					\end{prooftree}
				} s \text{ \fresh{} } out, pk_s
				\justifies
				\mbox{$\squeezespaces{1}\spec{UPD} \lts{\co{out}(pk_s)} \nw s.\mathopen{\left(\sub{pk_s}{\pks{s}}\right) \cpar \bang S}$}
				\using  \mbox{\textsf{Res}}
			\end{prooftree}
		}
		\]
		\captionsetup{labelformat=empty}
		\caption{\textit{Case} 1. Transition $\spec{UPD} \lts{\co{out}(pk_s)} \spec{UPD}^{\emptyset}(\emptyset)$.}
	\end{figure*}

	\begin{figure*}[h]
		\[
		\scalebox{1}{
			\begin{prooftree}
				{
					\begin{prooftree}
						pk_s \text{ \fresh{} } out, s, \bang I \quad out \mathrel{=_{E}} out
						\justifies
						\mbox{$\squeezespaces{1}\cout{out}{\pks{s}}.\bang I \lts{\co{out}(pk_s)} \mathopen{\left(\sub{pk_s}{\pks{s}}\right) \cpar \bang I}$}
						\using  \mbox{\textsf{Out}}
					\end{prooftree}
				} s \text{ \fresh{} } out, pk_s
				\justifies
				\mbox{$\squeezespaces{1}\impl{UPD} \lts{\co{out}(pk_s)} \nw s.\mathopen{\left(\sub{pk_s}{\pks{s}}\right) \cpar \bang I}$}
				\using  \mbox{\textsf{Res}}
			\end{prooftree}
		}
		\]
		\captionsetup{labelformat=empty}
		\caption{\textit{Case} 1. Transition $\impl{UPD} \lts{\co{out}(pk_s)} \impl{UPD}^{\emptyset,\emptyset}(\emptyset)$.}
	\end{figure*}
	
	\begin{figure*}[h]
		\[
		\scalebox{0.77}{
			\begin{prooftree}{
					\begin{prooftree}{
							\begin{prooftree}{
									\begin{prooftree}{
											\begin{prooftree}
												\begin{aligned} & u_{L+1} \text{ \fresh{} } card, ch, \fix{\card}(s, c_{L+1}, ch_{L+1}), \sigma \\ & card \sigma \mathrel{=_{E}} card \end{aligned}
												\justifies
												\begin{aligned}&\cout{card}{ch_{L+1}}.\fix{\card}(s, c_{L+1}, ch_{L+1}) \\ & \lts{\co{card}(u_{L+1})} \\ & \sigma \circ \sub{u_{L+1}}{ch_{L+1}} \cpar \m{E}^{L+1}(ch_{L+1})\end{aligned}
												\using \textsf{Out}
										\end{prooftree}} \begin{aligned}&c_{L+1}, ch_{L+1} \text{ \fresh{} } \\ & card, u_{L+1}, \sigma \end{aligned}
										\justifies
										\begin{aligned}&\sigma \cpar S \\ & \lts{\co{card}(u_{L+1})} \\& \nu c_{L+1}, ch_{L+1}.(\sigma \circ \sub{u_{L+1}}{ch_{L+1}} \cpar \m{E}^{L+1}(ch_{L+1}))\end{aligned}
										\using \textsf{Extrusion}^2
								\end{prooftree}} \begin{aligned} & c_{L+1}, ch_{L+1}, \\ & u_{L+1} \text{ \fresh{} } S \end{aligned}
								\justifies 
								\begin{aligned} & \sigma \cpar \bang S \\ & \lts{\co{card}(u_{L+1})} \\ & \nu c_{L+1}, ch_{L+1}.(\sigma \circ \sub{u_{L+1}}{ch_{L+1}} \cpar \m{E}^{L+1}(ch_{L+1}) \cpar \bang S) \end{aligned}
								\using \textsf{Rep-act}
						\end{prooftree}} \begin{aligned} & c_{L+1}, ch_{L+1}, u_{L+1} \text{ \fresh{} } \\ & C_i, i \leq L\end{aligned}
						\justifies
						\begin{aligned} & \sigma \cpar C_1 \cpar \hdots \cpar C_L \cpar \bang S \\ & \lts{\co{card}(u_{L+1})} \\ & \nu c_{L+1}, ch_{L+1}.(\sigma \circ \sub{u_{L+1}}{ch_{L+1}} \cpar C_1 \cpar \hdots \cpar C_L \cpar \m{E}^{L+1}(ch_{L+1}) \cpar \bang S) \end{aligned}
						\using \textsf{Par}^L
				\end{prooftree}} \begin{aligned} & s, c_i, ch_i, a_k \\ & i \leq L, k \in \beta \cup \gamma \cup \delta \text{ \fresh{} } \\ & card, u_{L+1} \end{aligned}
				\justifies
				\spec{UPD}^{\Psi}(\vec{Y}) \lts{\co{card}(u_{L+1})} \nu s, c_1, \hdots, c_L, c_{L+1}, ch_1, \hdots, ch_L, ch_{L+1}, a_{l_1}, \hdots, a_{l_K}.(\sigma \circ \sub{u_{L+1}}{ch_{L+1}} \cpar C_1 \cpar \hdots \cpar C_L \cpar \m{E}^{L+1}(ch_{L+1}) \cpar \bang S)
				\using \textsf{Res}^{1+2L+K}
			\end{prooftree}
		}
		\]
		\captionsetup{labelformat=empty}
		\caption{\textit{Case} 2. Transition $\spec{UPD}^{\Psi}(\vec{Y}) \lts{\co{card}(u_{L+1})} \spec{UPD}^{\{\alpha \cup \{L+1\}, \beta, \gamma, \delta\}}((Y_1, \hdots, Y_L, \emptyset))$.}
	\end{figure*}
	
	\begin{figure*}
		\[
		\scalebox{0.69}{
			\begin{prooftree}{
					\begin{prooftree}{
							\begin{prooftree}{
									\begin{prooftree}{
											\begin{prooftree}
												\begin{aligned} & u_{L+1} \text{ \fresh{} } card, ch_{L+1}, \fix{\card}(s, c_d, ch_{L+1}), \theta \\ & card \theta \mathrel{=_{E}} card \end{aligned}
												\justifies
												\begin{aligned}& \theta \cpar \cout{card}{ch_{L+1}}.\fix{C}(s, c_d, ch_{L+1}) \\ & \lts{\co{card}(u_{L+1})} \\ & \theta \circ \sub{u_{L+1}}{ch_{L+1}} \cpar \m{E}^d(ch_{L+1})\end{aligned}
												\using \textsf{Out}
										\end{prooftree}} \begin{aligned} & ch_{L+1} \textsf{ \fresh{} } \\ & card, u_{L+1}, \theta \end{aligned}
										\justifies
										\begin{aligned}& \theta \cpar \nw ch.\cout{card}{ch}.\fix{C}(s, c_d, ch) \\ & \lts{\co{card}(u_{L+1})} \\ & \nu ch_{L+1}.(\theta \circ \sub{u_{L+1}}{ch_{L+1}} \cpar \m{E}^d(ch_{L+1})\end{aligned}
										\using \textsf{Extrusion}
								\end{prooftree}} \begin{aligned} & ch_{L+1}, u_{L+1} \text{ \fresh{} } \\ & \nw ch.\cout{card}{ch}.\\ & \fix{C}(s, c_d, ch)\end{aligned}
								\justifies
								\begin{aligned}& \theta \cpar \bang \nw ch.\cout{card}{ch}.\fix{C}(s, c_d, ch) \\ & \lts{\co{card}(u_{L+1})} \\ & \nu ch_{L+1}.(\theta \circ \sub{u_{L+1}}{ch_{L+1}} \cpar \m{E}^d(ch_{L+1}) \cpar \bang \nw ch.\cout{card}{ch}.\fix{C}(s, c_d, ch))\end{aligned}
								\using \textsf{Rep-act}
						\end{prooftree}} \begin{aligned} & ch_{L+1}, u_{L+1} \text { \fresh{} } \\ & C^i_j, i \leq D, j \leq \max_{i \leq D}L_i; \\ & \nw ch.\cout{card}{ch}. \\ & \fix{C}(s, c_i, ch), \\ & i \leq D, i \neq d; \bang I \end{aligned}
						\justifies 
						\begin{aligned} & \theta \cpar \hdots \cpar \bang \nw ch.\cout{card}{ch}.\fix{C}(s, c_d, ch) \cpar \hdots \cpar \bang I\\ & \lts{\co{card}(u_{L+1})} \\ & \nu ch_{L+1}.(\theta \circ \sub{u_{L+1}}{ch_{L+1}} \cpar \hdots \cpar \m{E}^d(ch_{L+1}) \cpar \bang \nw ch.\cout{card}{ch}.\fix{C}(s, c_d, ch) \cpar \hdots \cpar \bang I)\end{aligned}
						\using \textsf{Par}^{D+L}
				\end{prooftree}} \begin{aligned} & s, c_i, ch_j, a_k, \\ & i \leq D, j \leq L, k \in \beta \cup \gamma \cup \delta \text{ \fresh{} } \\ & card, u_{L+1} \end{aligned}
				\justifies
				\impl{UPD}^{\Psi, \Omega}(\vec{Y}) \lts{\co{card}(u_{L+1})} \nu s, c_1, \hdots, c_D, ch_1, \hdots, ch_L, ch_{L+1}, a_{l_1}, \hdots, a_{l_K}.(\theta \circ \sub{u_{L+1}}{ch_{L+1}} \cpar \hdots \cpar \m{E}^d(ch_{L+1})\cpar \bang \nw ch.\cout{card}{ch}.\fix{C}(s, c_d, ch) \cpar \hdots \cpar \bang I)
				\using \textsf{Res}^{1+D+L+K}
			\end{prooftree}
		}
		\]
		\captionsetup{labelformat=empty}
		\caption{\textit{Case} 2. Transition $\impl{UPD}^{\Psi, \Omega}(\vec{Y}) \lts{\co{card}(u_{L+1})} \impl{UPD}^{\{\alpha \cup \{L+1\}, \beta, \gamma, \delta\}, \{\hdots, \zeta^d \cup \{L+1\} , \hdots\}}((Y_1, \hdots, Y_L, \emptyset))$: card $d$ starts new session.}
	\end{figure*}
	
	\begin{figure*}
		\[\scalebox{0.6}{
			\begin{prooftree}{
					\begin{prooftree}{
							\begin{prooftree}{
									\begin{prooftree}{
											\begin{prooftree}{
													\begin{prooftree}{
															\begin{prooftree}
																\begin{aligned} & u_{L+1} \text{ \fresh{} } card, ch_{L+1}, \fix{C}(s, c_{D+1}, ch_{L+1}) ,\theta \\ & card \theta \mathrel{=_{E}} card \end{aligned}
																\justifies
																\begin{aligned}& \theta \cpar \cout{card}{ch_{L+1}}.\fix{C}(s, c_{D+1}, ch_{L+1}) \\ & \lts{\co{card}(u_{L+1})} \\ & \theta \circ \sub{u_{L+1}}{ch_{L+1}} \cpar \m{E}^{D+1}(ch_{L+1}) \end{aligned}
																\using \textsf{Out}
														\end{prooftree}} \begin{aligned} & ch_{L+1} \text{ \fresh{} } \\ & card, \\ &  u_{L+1}, \theta \end{aligned}
														\justifies
														\begin{aligned}& \theta \cpar \nw ch.\cout{card}{ch}.\fix{C}(s, c_{D+1}, ch) \\ & \lts{\co{card}(u_{L+1})} \\ & \nw ch_{L+1}. (\theta \circ \sub{u_{L+1}}{ch_{L+1}} \cpar \m{E}^{D+1}(ch_{L+1}) \end{aligned}
														\using \textsf{Extrusion}
												\end{prooftree}} \begin{aligned} & ch_{L+1}, u_{L+1} \text{ \fresh{} } \\ & \nw ch.\cout{card}{ch}. \\ & \fix{C}(s, c_{D+1}, ch) \end{aligned}
												\justifies
												\begin{aligned}& \theta \cpar \bang \nw ch.\cout{card}{ch_{L+1}}.\fix{C}(s, c_{D+1}, ch) \\ & \lts{\co{card}(u_{L+1})} \\ & \nw ch_{L+1}. (\theta \circ \sub{u_{L+1}}{ch_{L+1}} \cpar \m{E}^{D+1}(ch_{L+1}) \cpar \bang \nw ch.\cout{card}{ch}.\fix{C}(s, c_{D+1}, ch)) \end{aligned}
												\using \textsf{Rep-act}
										\end{prooftree}} \begin{aligned} & c_{D+1} \textsf{ \fresh{} } \\ & card, \\ & u_{L+1}, \theta \end{aligned}
										\justifies
										\begin{aligned}& \theta \cpar I \\ & \lts{\co{card}(u_{L+1})} \\ & \nu c_{D+1}, ch_{L+1}.(\theta \circ \sub{u_{L+1}}{ch_{L+1}} \cpar \m{E}^{D+1}(ch_{L+1}) \cpar \bang \nw ch.\cout{card}{ch}.\fix{C}(s, c_{D+1}, ch))\end{aligned}
										\using \textsf{Extrusion}
								\end{prooftree}} \begin{aligned} & c_{D+1}, \\ & ch_{L+1}, \\ & u_{L+1} \text{ \fresh{} } I \end{aligned}
								\justifies
								\begin{aligned} & \theta \cpar \bang I\\ & \lts{\co{card}(u_{L+1})} \\ & \nu c_{D+1}, ch_{L+1}.(\theta \circ \sub{u_{L+1}}{ch_{L+1}} \cpar \m{E}^{D+1}(ch_{L+1}) \cpar \bang \nw ch.\cout{card}{ch}.\fix{C}(s, c_{D+1}, ch) \cpar \bang I)\end{aligned}
								\using \textsf{Rep-act}
						\end{prooftree}} \begin{aligned} & c_{D+1}, ch_{L+1}, \\ & u_{L+1} \text { \fresh{} } C^i_j, \\ & i \leq D,  j \leq \max_{i \leq D}L_i; \\ & \nw ch.\cout{card}{ch}. \\ & \fix{C}(s, c_d, ch) \end{aligned}
						\justifies 
						\begin{aligned} & \theta \cpar \hdots \cpar \bang I\\ & \lts{\co{card}(u_{L+1})} \\ & \nu c_{D+1}, ch_{L+1}.(\theta \circ \sub{u_{L+1}}{ch_{L+1}} \cpar \hdots \cpar \m{E}^{D+1}(ch_{L+1}) \cpar \bang \nw ch.\cout{card}{ch}.\fix{C}(s, c_{D+1}, ch) \cpar \bang I)\end{aligned}
						\using \textsf{Par}^{D+L}
				\end{prooftree}} \begin{aligned} & s, c_i, ch_j, a_k, \\ & i \leq D, j \leq L, \\ & k \in \beta \cup \gamma \cup \delta \text{ \fresh{} } \\ & card, u_{L+1} \end{aligned}
				\justifies
				\impl{UPD}^{\Psi, \Omega}(\vec{Y}) \lts{\co{card}(u_{L+1})} \nu s, c_1, \hdots, c_D, c_{D+1}, ch_1, \hdots, ch_L, ch_{L+1}, a_{l_1}, \hdots, a_{l_K}.(\theta \circ \sub{u_{L+1}}{ch_{L+1}} \cpar \hdots \cpar \m{E}^{D+1}(ch_{L+1}) \cpar \bang \nw ch.\cout{card}{ch}.\fix{C}(s, c_{D+1}, ch) \cpar \bang I)
				\using \textsf{Res}^{1+D+L+K}
			\end{prooftree}
		}
		\]
		\captionsetup{labelformat=empty}
		\caption{\textit{Case} 2. Transition $\impl{UPD}^{\Psi, \Omega}(\vec{Y}) \lts{\co{card}(u_{L+1})} \impl{UPD}^{{\{\alpha \cup \{L+1\}, \beta, \gamma, \delta\}}, \Omega \cup \{\{L+1\}\}}((Y_1, \hdots, Y_L, \emptyset))$: a new card is created.}
	\end{figure*}
	
	\begin{figure*}
		\[\scalebox{0.68}{
			\begin{prooftree}{
					\begin{prooftree}{
							\begin{prooftree}{
									\begin{prooftree}
										\begin{aligned} & v_l \text{ \fresh{} } u_l, a_l, \m{F}^{l}(ch_l, a_l), \sigma \\ &  u_l \sigma \mathrel{=_{E}} ch_l \end{aligned}
										\justifies
										\sigma \cpar \cout{ch_l}{\smult{a_l}{\pk{c_l}}}.\m{F}^{l}(ch_l, a_l) \lts{\co{u_l}(v_l)} \sigma \circ \sub{v_l}{\smult{a_l}{\pk{c_l}}} \cpar \m{F}^{l}(ch_l, a_l)
										\using \textsf{Out}
								\end{prooftree}} a_l \text{ \fresh{} } u_l, v_l, \sigma
								\justifies
								\sigma \cpar \nw a. \cout{ch_l}{\smult{a}{\pk{c_l}}}.\m{F}^l(ch_l, a) \lts{\co{u_l}(v_l)} \nu a_l.(\sigma \circ \sub{v_l}{\smult{a_l}{\pk{c_l}}} \cpar \m{F}^{l}(ch_l, a_l))
								\using \textsf{Extrusion}
						\end{prooftree}} a_l, v_l \text{ \fresh{} } C_i, i \leq L, i \neq l; \bang S 
						\justifies
						\sigma \cpar C_1 \cpar \hdots \cpar \m{E}^l(ch_l) \cpar \hdots \cpar C_L \cpar \bang S \lts{\co{u_l}(v_l)} \nu a_l.(\sigma \circ \sub{v_l}{\smult{a_l}{\pk{c_l}}} \cpar \hdots \cpar C_K \cpar \hdots \cpar \m{F}^{l}(ch_l, a_l) \cpar \hdots \cpar \bang S)
						\using \textsf{Par}^{L}
				\end{prooftree}} \begin{aligned} & s, c_i, ch_i, a_k \\ & i \leq L, k \in \beta \cup \gamma \cup \delta \text{ \fresh{} } u_l, v_l \end{aligned} 
				\justifies
				\spec{UPD}^{\Psi}(\vec{Y}) \lts{\co{u_l}(v_l)} \nu s, c_1, \hdots, c_L, ch_1, \hdots, ch_L, a_{l_1}, \hdots, a_{l_K}, a_l.(\sigma \circ \sub{v_l}{\smult{a_l}{\pk{c_l}}} \cpar \hdots \cpar C_K \cpar \hdots \cpar\m{F}^{l}(ch_l, a_l) \cpar \hdots \cpar \bang S)
				\using \textsf{Res}^{1+2L+K}
			\end{prooftree}
		}
		\]
		\captionsetup{labelformat=empty}
		\caption{\textit{Case} 3. Transition $\spec{UPD}^{\Psi}(\vec{Y}) \lts{\co{u_l}(v_l)} \spec{UPD}^{\{\alpha \setminus\{l\}, \beta \cup \{l\}, \gamma, \delta\}}(\vec{Y})$, $l \in \alpha$.}
	\end{figure*}
	
	\begin{figure*}
		\[\scalebox{0.68}{
			\begin{prooftree}{
					\begin{prooftree}{
							\begin{prooftree}{
									\begin{prooftree}
										\begin{aligned} & v_l \text{ \fresh{} } u_l, a_l, \m{F}^{d}(ch_l, a_l), \theta \\ &  u_l \theta \mathrel{=_{E}} ch_l \end{aligned}
										\justifies
										\theta \cpar \cout{ch_l}{\smult{a_l}{\pk{c_d}}}.\m{F}^{d}(ch_l, a_l) \lts{\co{u_l}(v_l)} \theta \circ \sub{v_l}{\smult{a_l}{\pk{c_d}}} \cpar\m{F}^{d}(ch_l, a_l)
										\using \textsf{Out}
								\end{prooftree}} a_l \text{ \fresh{} } u_l, v_l, \theta
								\justifies
								\theta \cpar \nw a. \cout{ch_l}{\smult{a}{\pk{c_d}}}.\m{F}^{d}(ch_l, a) \lts{\co{u_l}(v_l)} \nu a_l.(\theta \circ \sub{v_l}{\smult{a_l}{\pk{c_d}}} \cpar\m{F}^{d}(ch_l, a_l))
								\using \textsf{Extrusion}
						\end{prooftree}} \begin{aligned}& a_l, v_l \text{ \fresh{} } C^i_j, \\ & i \leq D,  j \leq \max_{i \leq D}L_i, \\ & j \neq l; \bang I \end{aligned}
						\justifies
						\theta \cpar \hdots \cpar \m{E}^d(ch_l) \cpar \hdots \cpar \bang I\lts{\co{u_l}(v_l)} \nu a_l.(\theta \circ \sub{v_l}{\smult{a_l}{\pk{c_d}}} \cpar \hdots \cpar\m{F}^{d}(ch_l, a_l) \cpar \hdots \cpar \bang I)
						\using \textsf{Par}^{D+L}
				\end{prooftree}} \begin{aligned} & s, c_i, ch_j, a_k \\ & i \leq D, j \leq L, k \in \beta \cup \gamma \cup \delta \text{ \fresh{} } \\ & u_l, v_l \end{aligned}
				\justifies
				\impl{UPD}^{\Psi, \Omega}(\vec{Y}) \lts{\co{u_l}(v_l)} \nu s, c_1, \hdots, c_D, ch_1, \hdots, ch_L, a_{l_1}, \hdots, a_{l_k}, a_l.(\theta \circ \sub{v_l}{\smult{a_l}{\pk{c_d}}} \cpar \hdots \cpar\m{F}^{d}(ch_l, a_l) \cpar \hdots \cpar \bang I)
				\using \textsf{Res}^{1+D+L+K}
			\end{prooftree}
		}
		\]
		\captionsetup{labelformat=empty}
		\caption{\textit{Case} 3. Transition $\impl{UPD}^{\Psi, \Omega}(\vec{Y}) \lts{\co{u_l}(v_l)} \impl{UPD}^{\alpha \setminus \{l\}, \beta \cup \{l\}, \gamma, \delta\}, \Omega}(\vec{Y})$, $l \in \alpha$.}
	\end{figure*}
	
	\begin{figure*}
		\[\scalebox{0.82}{
			\begin{prooftree}{
					\begin{prooftree}{
							\begin{prooftree}
								u_l \sigma \mathrel{=_{E}} ch_l
								\justifies
								\sigma \cpar \cin{ch_l}{y}. \m{G}^l(ch_l, a_l, y) \lts{u_l \, Y_l} \sigma \cpar \m{G}^l(ch_l, a_l, Y_l\sigma)
								\using \textsf{Inp}
						\end{prooftree}}
						\justifies
						\sigma \cpar C_1 \cpar \hdots \cpar \m{F}^l(ch_l, a_l) \cpar \hdots \cpar C_L \cpar \bang S  \lts{u_l \, Y_l} \sigma \cpar \hdots \cpar \m{G}^l(ch_l, a_l, Y_l\sigma) \cpar \hdots \cpar \bang S\
						\using \textsf{Par}^L
				\end{prooftree}} \begin{aligned} & s, c_i, ch_i, a_k \\ & i \leq L, k \in \beta \cup \gamma \cup \delta \text{ \fresh{} } u_l, Y_l \end{aligned} 
				\justifies
				\spec{UPD}^{\Psi}(\vec{Y}) \lts{u_l \, Y_l} \nu s, c_1, \hdots, c_L, ch_1, \hdots, ch_L, a_{l_1}, \hdots, a_{l_k}.\{\sigma \cpar \hdots \cpar \m{G}^l(ch_l, a_l, Y_l\sigma) \cpar \hdots \cpar \bang S\}
				\using \textsf{Res}^{1+2L+K}
			\end{prooftree}
		}
		\]
		\captionsetup{labelformat=empty}
		\caption{\textit{Case} 4. Transition $\spec{UPD}^{\Psi}(\vec{Y}) \lts{u_l \, Y_l} \spec{UPD}^{\{\alpha, \beta \setminus \{l\}, \gamma \cup \{l\}, \delta\}}(\chi_l(\vec{Y}, Y_l))$, $l \in \beta$.}
	\end{figure*}
	
	\begin{figure*}
		\[\scalebox{0.81}{
			\begin{prooftree}{
					\begin{prooftree}{
							\begin{prooftree}
								u_l \theta \mathrel{=_{E}} ch_l
								\justifies
								\theta \cpar \cin{ch_l}{y}. \m{G}^d(ch_l, a_l, y) \lts{u_l \, Y_l} \theta \cpar \m{G}^d(ch_l, a_l, Y_l\sigma)
								\using \textsf{Inp}
						\end{prooftree}}
						\justifies
						\theta \cpar \hdots \cpar \m{F}^d(ch_l, a_l) \cpar \hdots \cpar \bang I  \lts{u_l \, Y_l} \theta \cpar \hdots \cpar \m{G}^d(ch_l, a_l, Y_l\sigma) \cpar \hdots \cpar \bang I
						\using \textsf{Par}^{D+L}
				\end{prooftree}} \begin{aligned} & s, c_i, ch_j, a_k \\ & i \leq D, j \leq L, k \in \beta \cup \gamma \cup \delta \text{ \fresh{} } \\ & u_l, Y_l \end{aligned}
				\justifies
				\impl{UPD}^{\Psi, \Omega}(\vec{Y}) \lts{u_l \, Y_l} \nu s, c_1, \hdots, c_D, ch_1, \hdots, ch_L, a_{l_1}, \hdots, a_{l_K}.\{\theta \cpar \hdots \cpar \m{G}^d(ch_l, a_l, Y_l\sigma) \cpar \hdots \cpar \bang I\}
				\using \textsf{Res}^{1+D+L+K}
			\end{prooftree}
		}
		\]
		\captionsetup{labelformat=empty}
		\caption{\textit{Case} 4. Transition $\impl{UPD}^{\Psi, \Omega}(\vec{Y}) \lts{u_l \, Y_l} \impl{UPD}^{\{\alpha, \beta \setminus \{l\}, \gamma \cup \{l\}, \delta\}, \Omega}(\chi_l(\vec{Y}, Y_l))$ if there is a card at the stage $\m{F}$.}
	\end{figure*}
	
	\begin{figure*}
		\[\scalebox{0.83}{
			\begin{prooftree}{
					\begin{prooftree}{
							\begin{prooftree}
								\begin{aligned} & w_l \text{ \fresh{} } u_l, m^l(a_l, Y_l\sigma'), \sigma \\ & u_l \theta \mathrel{=_{E}} ch_l \end{aligned}
								\justifies
								\sigma \cpar \cout{ch_l}{m^l(a_l, Y_l\sigma')} \lts{\co{u_l}(w_l)} \sigma \circ \sub{w_l}{m^l(a_l, Y_l \sigma')} \cpar \m{H}^l
								\using \textsf{Out}
						\end{prooftree}}
						\justifies
						\sigma \cpar C_1 \cpar \hdots \cpar \m{G}^l(ch_l, a_l, Y_l\sigma') \cpar \hdots \cpar C_L \cpar \bang S \lts{\co{u_l}(w_l)} \sigma \circ \sub{w_l}{m^l(a_l, Y_l \sigma')} \cpar \hdots \cpar \m{H}^l \cpar \hdots \cpar \bang S\
						\using \textsf{Par}^{L}
				\end{prooftree}} \begin{aligned} & s, c_i, ch_i, a_k \\ & i \leq L, k \in \beta \cup \gamma \cup \delta \text{ \fresh{} } u_l, w_l \end{aligned}
				\justifies
				\spec{UPD}^{\Psi}(\vec{Y}) \lts{\co{u_l}(w_l)} \nu s, c_1, \hdots, c_L, ch_1, \hdots, ch_L, a_{l_1}, \hdots, a_{l_K}.\{\sigma \circ \sub{w_l}{m^l(a_l, Y_l \sigma')} \cpar \hdots \cpar \m{H}^l \cpar \hdots \cpar \bang S\}
				\using \textsf{Res}^{1+2L+K}
			\end{prooftree}
		}
		\]
		\captionsetup{labelformat=empty}
		\caption{\textit{Case} 5. Transition $\spec{UPD}^{\Psi}(\vec{Y}) \lts{\co{u_l}(w_l)} \spec{UPD}^{\{\alpha, \beta, \gamma \setminus \{l\}, \delta \cup \{l\}\}}(\vec{Y})$, $l \in \gamma$.}
	\end{figure*}
	
	\begin{figure*}
		\[\scalebox{0.83}{
			\begin{prooftree}{
					\begin{prooftree}{
							\begin{prooftree}
								\begin{aligned} & w_l \text{ \fresh{} } u_l, m^d(a_l, Y_l\theta'), \theta \\ & u_l \theta \mathrel{=_{E}} ch_l \end{aligned}
								\justifies
								\theta \cpar \cout{ch_l}{m^d(a_l, Y_l\theta')} \lts{\co{u_l}(w_l)} \theta \circ \sub{w_l}{m^d(a_l, Y_l \theta')} \cpar \m{H}^d
								\using \textsf{Out}
						\end{prooftree}}
						\justifies
						\theta \cpar \hdots \cpar \m{G}^d(ch_l, a_l, Y_l\theta') \cpar \hdots \cpar \bang I \lts{\co{u_l}(w_l)} \theta \circ \sub{w_l}{m^d(a_l, Y_l \theta')} \cpar \hdots \cpar \m{H}^d \cpar \hdots \cpar \bang I\
						\using \textsf{Par}^{D+L}
				\end{prooftree}} \begin{aligned} & s, c_i, ch_j, a_k \\ & i<\leq D, j \leq L, k \in \beta \cup \gamma \cup \delta \text{ \fresh{} } \\ & u_l, w_l \end{aligned}
				\justifies
				\impl{UPD}^{\Psi, \Omega}(\vec{Y}) \lts{\co{u_l}(w_l)} \nu s, c_1, \hdots, c_D, ch_1, \hdots, ch_L, a_{l_1}, \hdots, a_{l_K}.\{\theta \circ \sub{w_l}{m^d(a_l, Y_l \theta')} \cpar \hdots \cpar \m{H}^d \cpar \hdots \cpar \bang I\}
				\using \textsf{Res}^{1+D+L+K}
			\end{prooftree}
		}
		\]
		\captionsetup{labelformat=empty}
		\caption{\textit{Case} 5. Transition $\impl{UPD}^{\Psi, \Omega}(\vec{Y}) \lts{\co{u_l}(w_l)} \impl{UPD}^{\{\alpha, \beta, \gamma \setminus \{l\}, \delta \cup \{l\}\}, \Omega}(\vec{Y})$, $l \in \gamma$.}
		\vspace*{7in}	
	\end{figure*}

\end{document}